\documentclass[fleqn,usenatbib]{mnras}

\usepackage{newtxtext,newtxmath}
\usepackage[T1]{fontenc}

\usepackage{graphicx}
\usepackage{amsmath}
\usepackage{hyperref}
\usepackage{bm}
\usepackage[dvipsnames, table]{xcolor}
\usepackage{orcidlink}
\usepackage{float}

\newcommand{\Mpch}{\,h^{-1}\text{Mpc}}
\newcommand{\Gpch}{\,h^{-1}\text{Gpc}}
\newcommand{\Omegam}{\Omega_{\rm m}}
\newcommand{\Omegab}{\Omega_{\rm b}}
\newcommand{\omegac}{\omega_{\rm cdm}}
\newcommand{\omegab}{\omega_{\rm b}}

\newcommand{\Neff}{N_{\rm eff}}
\newcommand{\Msunh}{\,h^{-1}{\rm M_{\odot}}}
\newcommand{\Q}{{\rm Q}}
\newcommand{\nrun}{{\rm d}n_s/{\rm d}\ln k}

\usepackage{fontawesome}
\usepackage[colorinlistoftodos]{todonotes}

\title[SUNBIRD]{\texttt{SUNBIRD}: A simulation-based model for full-shape density-split clustering}

\author[]{\parbox{\textwidth}{
Carolina Cuesta-Lazaro$^{1, 2, 3}$\thanks{E-mail: cuestalz@mit.edu}\orcidlink{0000-0002-6069-2999},
Enrique Paillas$^{4, 5}$\orcidlink{0000-0002-4637-2868},
Sihan Yuan$^{7, 8, 9}$\orcidlink{0000-0002-5992-7586},
Yan-Chuan Cai$^{10}$\orcidlink{0000-0002-2128-866X},
Seshadri Nadathur$^{11}$\orcidlink{0000-0001-9070-3102},
Will J. Percival$^{4, 5, 6}$\orcidlink{0000-0002-0644-5727},
Florian Beutler$^{10}$\orcidlink{0000-0003-0467-5438},
Arnaud de Mattia$^{12}$\orcidlink{0000-0003-0920-2947},
Daniel Eisenstein$^{1}$\orcidlink{0000-0002-2929-3121},
Daniel Forero-Sanchez$^{13}$\orcidlink{0000-0001-5957-332X},
Nelson Padilla$^{14}$\orcidlink{0000-0001-9850-9419},
Mathilde Pinon$^{12}$\orcidlink{0009-0009-3228-7126},
Vanina Ruhlmann-Kleider$^{12}$\orcidlink{0009-0000-6063-6121},
Ariel~G.~S\'anchez$^{15}$\orcidlink{0000-0003-1198-831X},
Georgios Valogiannis$^{16,17}$\orcidlink{0000-0003-0805-1470},
and Pauline Zarrouk$^{18}$\orcidlink{0000-0002-7305-9578}
}
\vspace*{4pt} \\
\scriptsize $^{1}$ Center for Astrophysics | Harvard \& Smithsonian, 60 Garden St, Cambridge, MA 02138, USA  \\
\scriptsize $^{2}$ The NSF AI Institute for Artificial Intelligence and Fundamental Interactions \\
\scriptsize $^{3}$ Department of Physics, Massachusetts Institute of Technology, Cambridge, MA 02139, USA \\
\scriptsize $^{4}$ Waterloo Centre for Astrophysics, University of Waterloo, Waterloo, ON N2L 3G1, Canada \\
\scriptsize $^{5}$ Department of Physics and Astronomy, University of Waterloo, 
Waterloo, ON N2L 3G1, Canada \\
\scriptsize $^{6}$ Perimeter Institute for Theoretical Physics, 31 Caroline St North, Waterloo, ON N2L 2Y5, Canada \\
\scriptsize $^{7}$ Kavli Institute for Particle Astrophysics and Cosmology, Stanford University, 452 Lomita Mall, Stanford, CA 94305, USA\\
\scriptsize $^{8}$ Department of Physics, Stanford University, 382 Via Pueblo Mall, Stanford, CA 94305, USA\\
\scriptsize $^{9}$ SLAC National Accelerator Laboratory, 2575 Sand Hill Road, Menlo Park, CA  94025, USA \\
\scriptsize $^{10}$ Institute for Astronomy, University of Edinburgh, Blackford Hill, Edinburgh, EH9 3HJ, UK\\
\scriptsize $^{11}$ Institute of Cosmology and Gravitation, University of Portsmouth, Burnaby Road, Portsmouth, PO1 3FX, UK \\
\scriptsize $^{12}$ IRFU, CEA, Universit\'e Paris-Saclay, F-91191 Gif-sur-Yvette, France  \\
\scriptsize $^{13}$ Institute of Physics, Laboratory of Astrophysics, \'Ecole Polytechnique F\'ed\'erale de Lausanne (EPFL), Observatoire de Sauverny, CH-1290 Versoix, Switzerland \\
\scriptsize $^{14}$ Instituto de Astronom\'ia Te\'orica y Experimental (IATE), CONICET-Universidad Nacional de C\'ordoba, Laprida 854, X5000BGR, C\'ordoba, Argentina \\
\scriptsize $^{15}$ Max-Planck-Institut f\"ur Extraterrestrische Physik, Postfach 1312, Giessenbachstr., D-85748 Garching, Germany \\
\scriptsize $^{16}$ Department of Physics, Harvard University, Cambridge, MA, 02138, USA \\
\scriptsize $^{17}$ Department of Astronomy and Astrophysics, University of Chicago, Chicago, IL, 60637, USA \\
\scriptsize $^{18}$ Sorbonne Universit\'e, Universit\'e Paris Diderot, Sorbonne Paris Cit\'e, CNRS,
Laboratoire de Physique Nucl\'eaire et de Hautes Energies (LPNHE), 4 place Jussieu, F-75252, Paris
Cedex 5, France
\vspace*{-2pt}
}

\pubyear{2022}

\begin{document}

\label{firstpage}
\pagerange{\pageref{firstpage}--\pageref{lastpage}}
\maketitle

\begin{abstract}
Combining galaxy clustering information from regions of different environmental densities can help break cosmological parameter degeneracies and access non-Gaussian information from the density field that is not readily captured by the standard two-point correlation function (2PCF) analyses. However, modelling these density-dependent statistics down to the non-linear regime has so far remained challenging. We present a simulation-based model that is able to capture the cosmological dependence of the full shape of the density-split clustering (DSC) statistics down to intra-halo scales. Our models are based on neural-network emulators that are trained on high-fidelity mock galaxy catalogues within an extended-$\Lambda$CDM framework, incorporating the effects of redshift-space, Alcock-Paczynski distortions and models of the halo-galaxy connection. Our models reach sub-percent level accuracy down to $1 \Mpch$ and are robust against different choices of galaxy-halo connection modelling. When combined with the galaxy 2PCF, DSC can tighten the constraints on $\omegac$, $\sigma_8$, and $n_s$ by factors of 2.9, 1.9, and 2.1, respectively, compared to a 2PCF-only analysis. DSC additionally puts strong constraints on environment-based assembly bias parameters. Our code is made publicly available on Github \href{https://github.com/florpi/sunbird}{\faGithub}.
\end{abstract}


\section{Introduction}

The 3D clustering of galaxies contains a wealth of information about the contents and evolution of the Universe; from the properties of the early Universe to the nature of dark energy and dark matter, and to information on how galaxies form and evolve. Galaxy clustering provided some of the first evidence of the accelerated Universe \citep{10.1093/mnras/242.1.43P}, helped establish the standard model of cosmology through the detection of baryon acoustic oscillations \citep{Percival2001:astro-ph/0105252, Cole_2005, Eisenstein_2005}, and has yielded accurate cosmological constraints \citep{Anderson_2014}. Upcoming surveys such as DESI \citep{desi}, Euclid \citep{euclid}, and Roman \citep{roman} will probe unprecedented volumes, enabling more stringent constraints that may reveal inconsistencies challenging the standard cosmological model or our understanding of how galaxies form and evolve.

The spatial distribution of galaxies is commonly summarised by its two-point functions, the so-called two-point correlation function or its Fourier space equivalent, the power spectrum. For a Gaussian random field, this compression would be lossless. As the distribution of density fluctuations evolves through gravitational collapse, it becomes non-Gaussian: although overdensities can grow freely, underdensities are always bounded from below, as the density contrast in regions devoid of matter can never go below $\delta = -1$. As a consequence, the density field develops significant skewness and kurtosis, departing from Gaussianity \citep{Einasto2020}.

The induced non-Gaussianity in galaxy clustering deems the correlation function a lossy summary. For this reason, cosmologists have developed a wealth of summary statistics that may be able to extract more relevant information from the 3D clustering of galaxies. Examples include the three-point correlation function \citep{Slepian2017} or bispectrum \citep{Gil-Marin2017:1606.00439,Sugiyama2018:1803.02132v3, Philcox2021a}, the four-point correlation function \citep{Philcox2021b} or trispectrum \citep{Gualdi2021}, counts-in- cells statistics \citep{Szapudi2004, Klypin2018, Jamieson2020, Uhlemann2020}, non-linear transformations of the density field \citep{Neyrinck2009, Neyrinck2011b, Wang2011, Wang2022:1912.03392v3}, the separate universe approach \citep{Chiang2015}, the marked power spectrum \citep{Massara2018, Massara2022}, the wavelet scattering transform \citep{Valogiannis2021,Valogiannis_2022}, void statistics \citep{Hawken2020, Nadathur2020, Correa2020, Woodfinden2022}, k-nearest neighbours \citep{Banerjee:2020umh, 2023Yuan}, and other related statistics. Alternatively, one could avoid the use of summary statistics completely and attempt to perform inference at the field level \citep{lavaux2019systematicfree,Schmidt_2021, dai2023multiscale,Dai:2022dso}.

However, utilising these summary statistics has been limited by our inability to model them analytically over a wide range of scales, difficulty compressing their high dimensionality, or due to a lack of accurate perturbation theory predictions or the difficulty in modelling the effect that observational systematics have on arbitrary summary statistics \citep{Yuan_2023}. This has now drastically changed due to i) \textit{advancements in simulations}: we now can run large suites of high-resolution simulations in cosmological volumes \cite{DeRose_2019,Nishimichi_2019,Maksimova_2021}, which enable us to forward model the relation between the cosmological parameters and the summary statistics with greater accuracy; and ii) \textit{progress in machine learning techniques} that allow us to perform inference on any set of parameters, $\theta$, given any summary statistic, $s$, provided we can forward model the relation $s(\theta)$ for a small set of $\theta$ values \citep{Cranmer_2020}. Examples of the latter in cosmology are emulators, that model $s(\theta)$ mainly through neural networks or Gaussian processes \citep{Heitmann_2009, DeRose_2019, Zhai_2023} and assume a Gaussian likelihood, or density estimators used to model directly the posterior distribution $p(\theta|s(x))$ \citep{Jeffrey_2020, hahn2022rm} and make no assumptions about the likelihood's distribution.

While these advancements allow us to constrain cosmology with remarkable accuracy, our primary focus extends beyond just finding the most informative summary statistics. We are interested in statistics that could lead to surprising results revising our understanding of how the Universe formed and evolved. Notably, models beyond Einstein gravity that add degrees of freedom in the gravitational sector must screen themselves from local tests of gravity, and can therefore only deviate from general relativity in regions of low density or low gravitational potential \citep{Joyce_2015,2023Univ....9..302H}. Therefore, surprises in this direction could be found in statistics that explore the dependency of galaxy clustering to different density environments. Moreover, previous work \citep{Paillas2021, Paillas2022:2209.04310, Bonnaire2022} has demonstrated that these statistics also have a large constraining power on the cosmological parameters.

Although we have mentioned above that we can now run large suites of simulations in cosmological volumes, this is only true for N-body, dark matter-only simulations. We still need a flexible and robust galaxy-dark matter connection model that allows us to populate dark matter simulations with realistic galaxy distributions. In this work, we employ halo occupation distribution (HOD) models, which use empirical relations to describe the distribution of galaxies in a halo based on the halo's mass and other secondary halo properties. In particular, recent studies have found the halo local density to be a good tracer of dark matter halo secondary properties, both in hydrodynamical simulations \citep{2020MNRAS.493.5506H} and semi-analytical models of galaxy formation \citep{2021MNRAS.502.3242X}.

Here we present a full-shape theory model for galaxy clustering in different density environments that can be used to infer the cosmological parameters from observations in a robust manner. In a companion paper \citep{Paillas2023:2309.16541} we present the first cosmological constraints resulting from density-split clustering using the model presented in this manuscript that we apply to the BOSS DR12 CMASS data \citep{Dawson_2016, reid2015sdssiii}.

The paper is organised as follows. We define the observables and how we model them in Section~\ref{sec:simulation_based}. In Section~\ref{sec:validate_emulator}, we demonstrate that the model can accurately recover the parameters of interest in a range of mock galaxy obseravtions. We discuss our results and compare them to previous findings in the literature in Section~\ref{sec:discussion}.

\section{A simulation-based model for density split statistics}
\label{sec:simulation_based}

We are interested in modelling the connection between the cosmological parameters, $\mathcal{C}$, the additional parameters describing how galaxies populate the cosmic web of dark matter, $\mathcal{G}$, and clustering as a function of density environment, $\mathbf{X}^{\rm obs}$. To solve the inverse problem and consrain $\mathcal{C}$ and $\mathcal{G}$ from data, we could use simulated samples drawn from the joint distribution $p(\mathcal{C}, \mathcal{G}, \mathbf{X}^{\rm obs})$ to either, i) model the likelihood of the observation $p(\mathbf{X}^{\rm obs}|\mathcal{C}, \mathcal{G})$, subsequently sampling its posterior using Monte Carlo methods, or ii) directly model the posterior distribution $p(\mathcal{C}, \mathcal{G}|\mathbf{X}^{\rm obs})$, as demonstrated in \cite{Jeffrey_2020, hahn2022rm}, thus circumventing assumptions about the likelihood's functional form. Due to the Central Limit Theorem, we ancitipate the likelihood of galaxy pair counts to approximate a Gaussian distribution. In this section, we validate that this holds true specifically for density-split statistics and elucidate how simulations can model its mean and covaraince. Additionally, modelling the likelihood implies that we can use it as a measure of goodness of fit, vary the priors of the analysis at will, and combine our constraints with those of other independent observables.

In this section, we will proceed as follows: we begin by detailing our method for estimating density-dependent clustering. Subsequently, we discuss our approach for simulating the observable for a CMASS-like mock galaxy sample. We conclude by introducing our neural network model of the observable's likelihood.

\subsection{The observables}

\subsubsection{Two-point clustering}
\begin{figure*}
    \centering
    \begin{tabular}{ccc}
      \includegraphics[width=0.3\textwidth]{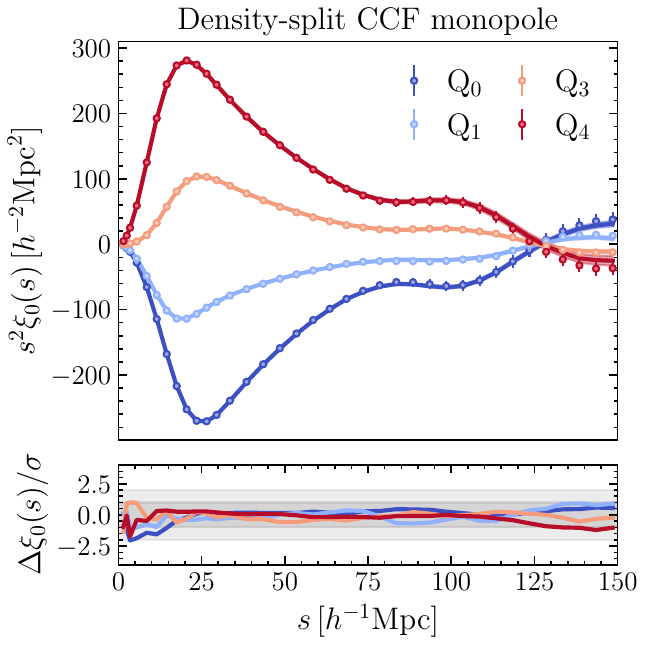}  & \includegraphics[width=0.3\textwidth]
      {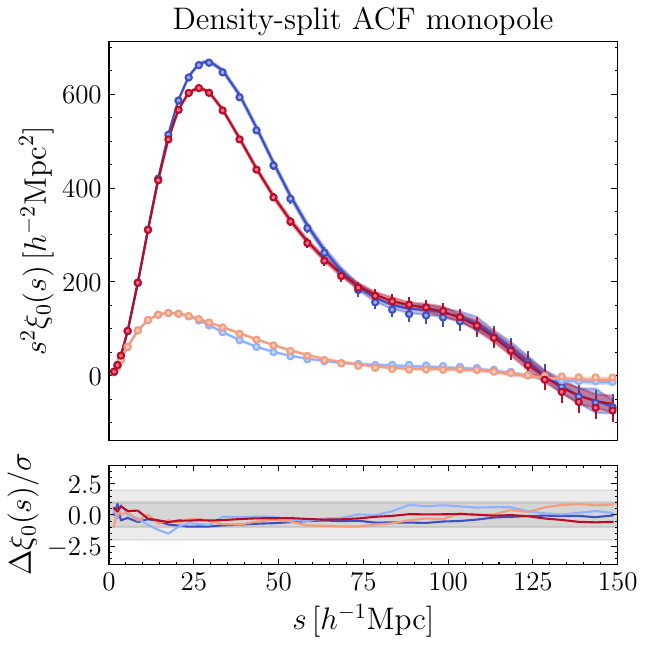} & \includegraphics[width=0.3\textwidth]{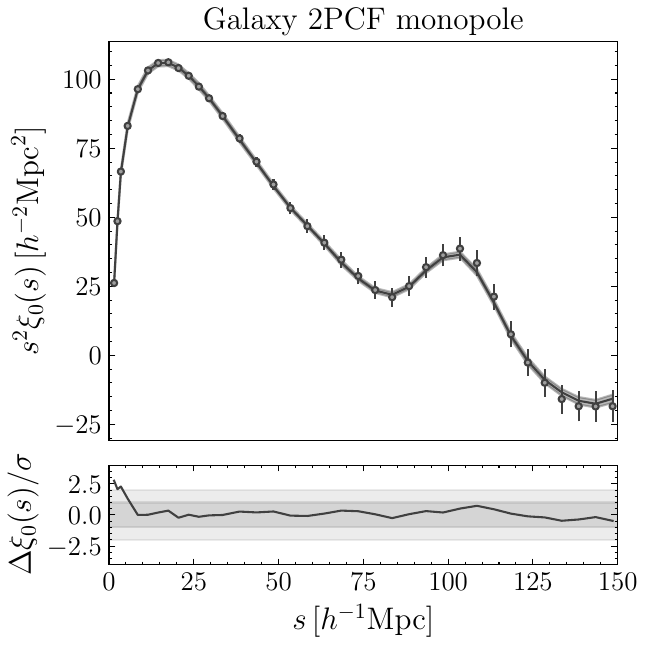}\\[4pt]
      \includegraphics[width=0.3\textwidth]{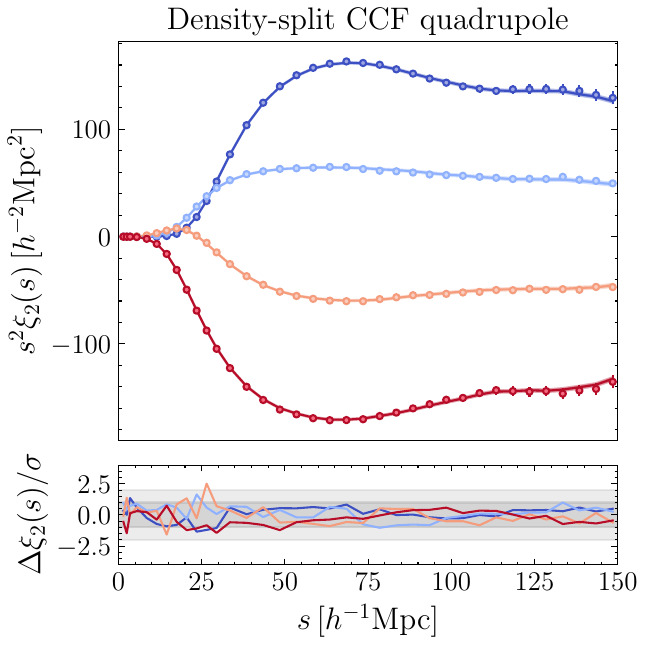}  & \includegraphics[width=0.3\textwidth]
      {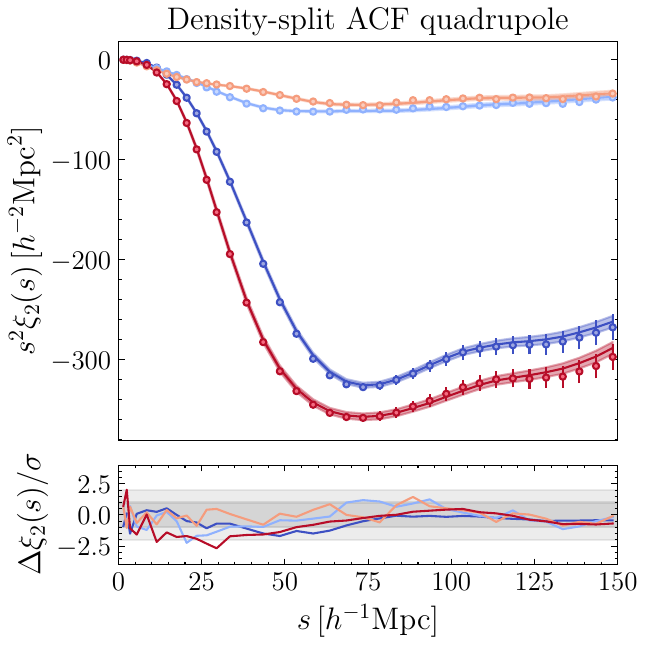} & \includegraphics[width=0.3\textwidth]{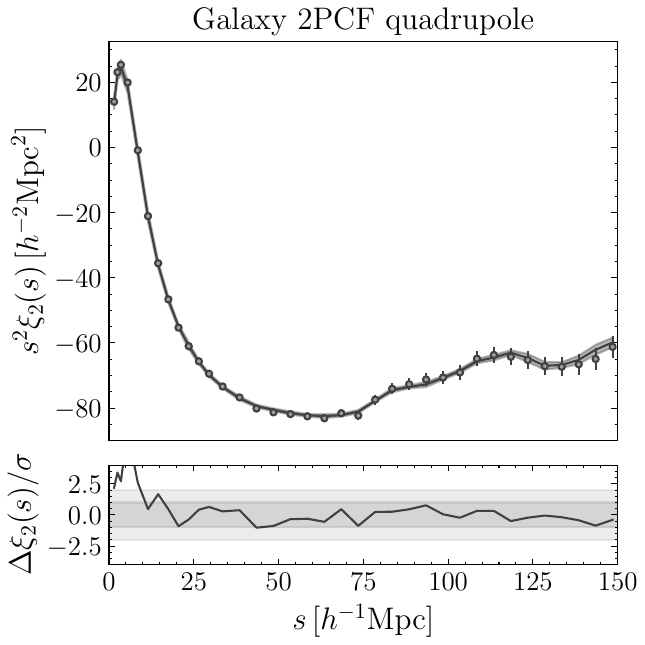}
    \end{tabular}
    \caption{A visualisation of the density-split clustering data vectors from the \textsc{AbacusSummit} simulations, along with emulator prediction at the parameter values of the simulation. The lowest density quintile is shown in blue, $\rm Q_0$, and the highest one in red, $\rm Q_4$. Markers and solid lines show the data vectors and the emulator predictions, respectively, whereas the shaded area represents the emulator predicted uncertainty. Left: multipoles of the quintile-galaxy cross-correlation functions. Middle: multipoles of the quintile autocorrelation functions. Right: multipoles of the two-point correlation function. The upper and lower panels show the monopole and quadrupole moments, respectively. We also display the difference between the model and the data, in units of the data error. Each colour corresponds to a different density quintile. \href{https://github.com/florpi/sunbird/blob/main/paper_figures/emulator_paper/F1_data_vectors.py}{\faGithub}}
    \label{fig:multipoles}
\end{figure*}
The information contained on 3D galaxy maps is commonly summarised in terms of the two-point correlation function (2PCF) $\xi^{\rm gg}(r)$(or the power spectrum in Fourier space), which measures the excess probability ${\rm d}P$ of finding a pair of galaxies separated by a scale $\bf{r}$ within a volume ${\rm d}V$, relative to an unclustered Poisson distribution,
\begin{equation}
    {\rm d}P = \overline{n} \left[1 + \xi^{\rm gg}(\bf{r}) \right]{\rm d}V \,,
\end{equation}
where $\overline{n}$ denotes the mean galaxy density. While the spatial distribution of galaxies is isotropic in real space, there are two main sources of distortions that induce anisotropies in the clustering measured from galaxy surveys: redshift-space distortions (RSD) and Alcock-Paczynski (AP) distortions, which are dynamical and geometrical in nature, respectively.

Redshift-space distortions arise when converting galaxy redshifts to distances ignoring the peculiar motion of the galaxies. A pair of galaxies that is separated by a vector ${\bf r}$ in real space, will instead appear separated by a vector ${\bf s}$ in redshift space (to linear order in velocity):
\begin{equation} \label{eq:RSD}
    {\bf s} = {\bf r} + \frac{{\bf v} \cdot \hat{\bf x}}{a(z) H(z)} \hat{\bf{x}}\, ,
\end{equation}
where $\hat{\bf x}$ is the unit vector associated with the observer's line of sight, ${\bf v}$ is the peculiar velocity of the galaxy, $a(z)$ is the scale factor and $H(z)$ is the Hubble parameter.\\

Alcock-Paczynski distortions arise when the cosmology that is adopted to convert angles and redshifts to distances, denoted as fiducial cosmology, differs from the true cosmology of the Universe. This effect is partially degenerate with RSD. For close pairs, the true pair separation is related to the observed pair separation via the parameters $q_{\perp}$ and $q_{\parallel}$, which distort the components of the pair separation across and along the observer's line of sight,
\begin{align}
    r_{\perp} = q_{\perp} r_{\perp}^{\rm fid}\, ; \mspace{10mu} r_{\parallel} = q_{\parallel} r_{\parallel}^{\rm fid} \ ,
\end{align}
where the $^{\rm fid}$ superscript represents the separations measured in the fiducial cosmology. The distortion parameters are given by
\begin{align} \label{eq:AP}
    q_{\parallel} = \frac{D_{\rm H}(z)}{D_{\rm H}^{\rm fid}(z)}\, ; \mspace{10mu} q_{\perp} = \frac{D_{\rm M}(z)}{D_{\rm M}^{\rm fid}(z)}\,,
\end{align}
where $D_{\rm M}(z)$ and $D_{\rm H}(z)$ are the comoving angular diameter and Hubble distances to redshift $z$, respectively.

Due to RSD and AP, the 2PCF is no longer isotropic but depends on $s$, the pair separation, and $\mu$, the cosine of the angle between the galaxy pair separation vector and the mid-point line of sight. The two-dimensional correlation function can be decomposed in a series of multipole moments,
\begin{equation}
    \label{eq:multipoles}
    \xi_\ell (s) = \frac{2 \ell + 1}{2} \int_{-1}^1 \rm d \mu \, \xi(s, \mu) P_\ell (\mu),
\end{equation}
where $\rm P_\ell$ is the $\ell$-th order Legendre polynomial.

\subsubsection{Density-split clustering}

The density-split method \citep{Paillas2022:2209.04310} characterises galaxy clustering in environments of different local densities. Instead of calculating the two-point clustering of the whole galaxy sample at once, one first splits a collection of randomly placed query points in different bins or `quantiles', according to the local galaxy overdensity at their locations. The two-point clustering is then calculated for each environment separately, and all this information is then combined in a joint likelihood analysis. The algorithm can be summarised as follows:

\begin{enumerate}
    \item Redshift-space galaxy positions are assigned to a rectangular grid with a cell size $R_{\rm cell}$, and the overdensity field is estimated using a cloud-in-cell interpolation scheme. The field is smoothed using a Gaussian filter with radius $R_s$, which is performed in Fourier space for computational efficiency.
    
    \item A set of $N_{\rm query}$ random points are divided into $N_Q$ density bins, or \textit{quantiles}, according to the overdensity measured at each point.

    \item Two summary statistics are calculated for each quantile: the autocorrelation function (DS ACF)  of the query points in each quantile, and the cross-correlation function (DS CCF) between the quantiles and the entire redshift-space galaxy field. These correlation functions are then decomposed into multipoles (equation~\ref{eq:multipoles}).
    
    \item The collection of correlation functions of all quantiles but the middle one, is combined in a joint data vector, which is then fitted in a likelihood analysis to extract cosmological information.
\end{enumerate}

In Fig.~\ref{fig:multipoles}, we show the different density split summary statistics for five quantiles and $R_s = 10\,h^{-1}{\rm Mpc}$, as measured in the \textsc{AbacusSummit} simulations presented in Section~\ref{sec:abacus_summit}. Note that the smoothing scale can be varied depending on the average density of tracers in a given survey, here we restrict ourselves to a smoothing scale appropriate for a CMASS-like survey. In the first column, we show the CCF of the different density quantiles and the entire galaxy sample. Above, the amplitude of the different correlations reflects the non-Gaussian nature of the density PDF:  the most underdense regions, $\rm Q_0$, are always constrained from below as voids cannot be emptier than empty ($\delta = -1$), meanwhile, dense regions, $\rm Q_4$, can go well beyond 1, breaking the symmetry of the correlations. Around the scale of $100 \Mpch$ we can distinguish the signal coming from the baryon acoustic oscillations for all density quantiles, both for the cross- and auto-correlations. Regarding the quadrupole moments, the anisotropy found is a consequence of the RSD effect on the galaxy positions, which also introduces an additional anisotropy in the distribution of quantiles when these are identified using the galaxy redshift space distribution, as shown in \citep{Paillas2022:2209.04310}.

\subsection{Forward modelling the galaxy observables}

In this subsection, we will first present the suite of dark-matter-only N-body simulations used in this work to model the cosmological dependence of density-split clustering, and will later present the galaxy-halo connection model we adopt to build CMASS-like mock galaxy catalogues. 

\subsubsection{The AbacusSummit simulations}
\label{sec:abacus_summit}

\textsc{AbacusSummit} \citep{Maksimova_2021} is a suite of cosmological N-body simulations that were run with the \textsc{Abacus} N-body code \citep{Garrison2019:1810.02916, Garrison2021:2110.11392}, designed to meet and exceed the simulation requirements of DESI \citep{Levi2019}. The base simulations follow the evolution of $6912^3$ dark matter particles in a $(2\,h^{-1}{\rm Gpc})^3$ volume, corresponding to a mass resolution of $2 \times 10^9\,{\rm M_{\odot}}/h$. 

In total, the suite spans $97$ different cosmologies, with varying,
\begin{align} \label{eq:cosmo_parameters}
    \mathcal{C} =  \{&\omega_{\rm cdm}, \omega_{b}, \sigma_8, n_s, \nrun, \Neff, w_0, w_a \},
\end{align}

where $\omegac = \Omega_{\rm c}h^2$ and $\Omega_{\rm b}h^2$ are the physical cold dark matter and baryon densities, $\nrun$ is the running of the spectral tilt, $\Neff$ is the effective number of ultra-relativistic species, $w_0$ is the present-day dark energy equation of state, and $w_a$ captures the time evolution of the dark energy equation of state. The simulations assume a flat spatial curvature, and the Hubble constant $H_0$ is calibrated to match the Cosmic Microwave Background acoustic scale $\theta_*$ to the Planck2018 measurement. 

In this study, we focus on the following subsets of the \textsc{AbacusSummit} simulations:
\begin{itemize}
    \item[] \fbox{\texttt{c000}} Planck2018 $\Lambda$CDM base cosmology \citep{Planck2020}, corresponding to the mean of the base\_plikHM\_TTTEEE\_lowl\_lowE\_lensing likelihood. There are $25$ independent realizations of this cosmology.

    \item[] \fbox{\texttt{c001-004}} Secondary cosmologies, including a low $\omega_{\rm cdm}$ choice \citep[WMAP7,][]{WMAP7}, a $w$CDM choice, a high $\Neff$ choice, and a low $\sigma_8$ choice.

    \item[] \fbox{\texttt{c013}} Cosmology that matches Euclid Flagship2 $
    \Lambda$CDM (Castander et al., in preparation).
    
    \item[] \fbox{\texttt{c100-126}} A linear derivative grid that provides pairs of simulations with small negative and positive steps in an 8-dimensional cosmological parameter space

    \item[] \fbox{\texttt{c130-181}} An emulator grid around the base cosmology that provides a wider coverage of the cosmological parameter space. Note that all the simulations in the emulator grid have the same phase seed. The parameter ranges in the emulator grid are shown in Table~\ref{tab:prior_emulator}.

\end{itemize}

\begin{table*}
    \centering
    \rowcolors{2}{white}{gray!15}
    \begin{tabular}{l l c r}
        \hline
        & Parameter & Interpretation & Prior range \\
        \hline
        Cosmology & $\omega_{\rm cdm}$ & Physical cold dark matter density & [0.103, 0.140] \\
        & $\omega_{\rm b}$ &  Physical baryon density & [0.0207, 0.024] \\
        & $\sigma_8$ & Amplitude of matter fluctuations in $8\,h^{-1}{\rm Mpc}$ spheres & [0.687, 0.938]\\
        & $n_s$ & Spectral index of the primordial power spectrum & [0.901, 1.025]\\
        & $\nrun$ & Running of the spectral index & [-0.038, 0.038]\\
        & $\Neff$ & Number of ultra-relativistic species & [2.1902, 3.9022]\\
        & $w_0$ & Present-day dark energy equation of state & [-1.27, -0.70]\\
        & $w_a$ & Time evolution of the dark energy equation of state & [-0.628, 0.621]\\
        \hline
        HOD & $M_{\rm cut}$ & Minimum halo mass to host a central & [12.4, 13.3]\\
        & $M_1$ & Typical halo mass to host one satellite & [13.2, 14.4]\\
        & $\log \sigma$ & Slope of the transition from hosting zero to one central & [-3.0, 0.0]\\
        & $\alpha$ & Power-law index for the mass dependence of the number of satellites & [0.7, 1.5]\\
        & $\kappa$ & Parameter that modulates the minimum halo mass to host a satellite & [0.0, 1.5]\\
        & $\alpha_c$ & Velocity bias for centrals & [0.0, 0.5]\\
        & $\alpha_s$ & Velocity bias for satellites & [0.7, 1.3]\\
        & $B_{\rm cen}$ & Environment-based assembly bias for centrals & [-0.5 0.5]\\
        & $B_{\rm sat}$ & Environment-based assembly bias for satellites & [-1.0, 1.0] \\
        \hline
    \end{tabular}
    \label{tab:prior_emulator}
    \caption{Definitions and ranges of the cosmological and galaxy-halo connection parameters for the simulations used to train our emulator.}

\end{table*}

Moreover, we use a smaller set of $1643$ N-body simulations denoted as \textsc{AbacusSmall} to estimate covariance matrices. These simulations are run with the same mass resolution as that of \textsc{AbacusSummit} in $500\,h^{-1}{\rm Mpc}$ boxes, with $1728^3$ particles and varying phase seeds.

Group finding is done on the fly, using a hybrid Friends-of-Friends/Spherical Overdensity algorithm, dubbed CompaSO \citep{Hadzhiyska2021:2110.11408}. We use dark matter halo catalogues from snapshots of the simulations at $z = 0.5$ and populate them with galaxies using the extended Halo Occupation Distribution framework presented in Sect.~\ref{subsubsec:hod}.

\subsubsection{Modelling the galaxy-halo connection}
\label{subsubsec:hod}

We model how galaxies populate the cosmic web of dark matter using the Halo Occupation Distribution (HOD) framework, which populates dark matter haloes with galaxies in a probabilistic way, assuming that the expected number of galaxies in each halo correlates with some set of halo properties, the main one being halo mass.

In the base halo model \citep{Zheng2007}, the average number of central galaxies in a halo of mass $M$ is given by
\begin{align}
    \langle N_{\rm c} \rangle(M) = \frac{1}{2} \left(1 + \mathrm{erf} \left(\frac{\log M - \log M_{\rm cut}}{\sqrt{2}\sigma} \right)  \right)\,,
\end{align}
where $\mathrm{erf}(x)$ denotes the error function, $M_{\rm cut}$ is the minimum mass required to host a central, and $\sigma$ is the slope of the transition between having zero to one central galaxy. The average number of satellite galaxies is given by
\begin{align}
    \langle N_{\rm s} \rangle(M) = \langle N_{\rm c} \rangle(M) \left(\frac{M - \kappa M_{\rm cut}}{M_1} \right)^{\alpha}\,
\end{align}
where $\kappa M_{\rm cut}$ gives the minimum mass required to host a satellite, $M_1$ is the typical mass that hosts one satellite, and $\alpha$ is the power law index for the number of galaxies. Note that these particular functional forms have been developed for the clustering of luminous red galaxies (LRGs) and should be modified for other tracers such as emission line galaxies (ELGs).

Alternatively, one could model the connection between dark matter halos and galaxies through more complex models of galaxy formation such as semi-analytical models or hydrodynamical simulations. In these scenarios, the simplified assumptions of HOD models whose occupation parameters solely depend on halo mass have been found to break down. In particular, recent studies have found the halo local density to be a good tracer of dark matter halo secondary properties that control galaxy occupation, both in hydrodynamical simulations \citep{2020MNRAS.493.5506H} and semi-analytical models of galaxy formation \citep{2021MNRAS.502.3242X}. There is however no direct observational evidence of this effect so far, and we are interested in using density split statistics to more accurately constrain the role that environment plays in defining the halo-galaxy connection.

In this work, we implement the HOD modelling using \textsc{AbacusHOD} \citep{Yuan2021:2110.11412}, which is a highly efficient \textsc{Python} package that contains a wide range of HOD variations. In \textsc{AbacusHOD}, the environment-based secondary bias parameters, $B_{\rm cen}$ \& $B_{\rm sat}$, effectively modulate the mass of a dark matter halo during the HOD assignment, so that it depends on the local matter overdensity $\delta_m$ 

\begin{alignat}{2}
    &\log_{10}M_{\rm cut}^{\rm eff} &&= \log_{10}M_{\rm cut} + B_{\rm cen} (\delta_m - 0.5) \nonumber \\
    &\log_{10}M_{1}^{\rm eff} &&= \log_{10}M_{1} + B_{\rm sat} (\delta_m - 0.5)\,.
\end{alignat}
Here, $\delta_m$ is defined as the mass density within a $5 \Mpch$ tophat filter from the halo centre, without considering the halo itself. More details about the exact implementation of this extension can be found in \cite{Yuan2021:2110.11412}.

Moreover, we include velocity bias parameters to increase the flexibility of the model to describe the dynamics of galaxies within dark matter haloes, that ultimately influence galaxy clustering through redshift-space distortions. There is in fact observational evidence pointing towards central galaxies having a larger velocity dispersion than their host dark matter halos \citep{Guo_2014, Yuan2021:2110.11412} for CMASS galaxies (dominated by LRGs), evidence for other tracers is not established yet. In the \textsc{AbacusHOD} implementation, the positions and velocities of central galaxies are matched to the most-bound particle in the halo, whereas the satellites follow the positions and velocities of randomly selected dark matter particles within the halo. The velocity bias parameters, $\alpha_{\rm vel, c}$ \& $\alpha_{\rm vel, s}$, allow for offsets in these velocities, such that the centrals do not perfectly track the velocity of the halo centre, and the satellites do not exactly match the dark matter particle velocities. The exact velocity match is recovered when $\alpha_{\rm vel, c} = 0$ and $\alpha_{\rm vel, c} = 1$.

The extended-HOD framework used in this study is then comprised of 9 parameters:
\begin{align} \label{eq:hod_parameters}
    \mathcal{G} = \{ M_{\rm cut}, M_1, \sigma, \alpha, \kappa, \alpha_{\rm vel, c}, \alpha_{\rm vel, s}, B_{\rm cen}, B_{\rm sat} \} \,.
\end{align}

Note that we are here not including additional parameters that may help marginalize over the effect that baryons have on halo density profiles. Although this has been shown to be a small effect \citep{Bose_2019}, \cite{Yuan2021:2110.11412} presented an extended parametrisation that could be use to marginalize over this effect. 

\subsubsection{Generating mock galaxy catalogues}

We generate a Latin hypercube with $8500$ samples from the 9-dimensional HOD parameter space defined in equation~\ref{eq:hod_parameters}, with parameter ranges as listed in Table~\ref{tab:prior_emulator}. Each of the $85$ cosmologies is assigned $100$ HOD variations from the Latin hypercube, which are then used to generate mock galaxy catalogues using the \textsc{AbacusHOD}. This number of HOD variations was chosen as a compromise between reducing the emulator error and increasing the computational cost of these measurements. In the future, we plan to develop a more efficient HOD sampling strategy to re-sample those HOD parameter values where the emulator error is large.

Our target galaxy sample is the DR12 BOSS CMASS galaxy sample \citep{reid2015sdssiii} at $0.45 < z < 0.6$. If the resulting number density of an HOD catalogue is larger than the observed number density from CMASS, $n_{\rm gal} \approx 3.5 \times 10^{-4}\,(h/{\rm Mpc})^{-3}$, we invoke an incompleteness parameter $f_{\rm ic}$ and randomly downsample the catalogue to match the target number density.

The resulting HOD catalogues consist of the real-space galaxy positions and velocities. Under the distant-observer approximation, we map the positions of galaxies to redshift space by perturbing their coordinates along the line of sight with their peculiar velocities along the same direction (equation~\ref{eq:RSD}). For each mock catalogue, we build three redshift-space counterparts by adopting three different lines of sight, taken to be the $x$, $y$ and $z$ axes of the simulation, which can be averaged out in the clustering analysis to increase the signal-to-noise ratio of the correlation functions \citep{Smith_2020}.

Since the end goal of our emulator is to be able to model galaxy clustering from observations, we adopt the same fiducial cosmology as in our CMASS clustering measurements \citep{Paillas2023:2309.16541},
\begin{align}
    &\omegac = 0.12 \quad \omegab = 0.02237 \quad
    h = 0.6736 \nonumber \\
    &\sigma_8 = 0.807952 \quad n_s= 0.9649\, ,
\end{align}
and infuse the mocks with the Alcock-Paczynski distortions that would be produced if we were to analyse each mock with this choice of fiducial cosmology. We do so by scaling the galaxy positions\footnote{These distortions would have been naturally produced if we had started from galaxy catalogues in sky coordinates, and used our fiducial cosmology to convert them to comoving cartesian coordinates. In our case, we have to manually distort the galaxy positions, since we are already starting from the comoving box.} and the simulation box dimensions with the distortion parameters from equation~\ref{eq:AP}, which depend on the adopted fiducial cosmology and the true cosmology of each simulation.
Since, in general, $q_{\perp}$ and $q_{\parallel}$ can be different, the box geometry can become non-cubic, but it still maintains the periodicity along the different axes. This is taken into account when calculating the clustering statistics, as explained in the next section.

\subsubsection{Generating the training sample}

We run the density-split clustering pipeline on the HOD mocks using our publicly available code\footnote{\href{https://github.com/epaillas/densitysplit}{https://github.com/epaillas/densitysplit}.}. Redshift-space galaxy positions are mapped onto a rectangular grid of resolution $R_{\rm cell} = 5\,h^{-1}{\rm Mpc}$, smoothed with a Gaussian kernel of width $R_s = 10\,h^{-1}{\rm Mpc}$. The overdensity field\footnote{The galaxy overdensity in each grid cell depends on the number of galaxies in the cell, the average galaxy number density, and the total number of grid cells. As we are working with a rectangular box with periodic boundary conditions, the average galaxy number density can be calculated analytically, which allows us to convert the galaxy number counts in each cell to an overdensity. When working with galaxy surveys, this has to be calculated using random catalogues that match the survey window function.} is sampled at $N_{\rm query}$ random locations, where $N_{\rm query}$ is equal to five times the number of galaxies in the box. We split the query positions into five quantiles according to the overdensity at each location. We plan to explore the constraining power of the statistic based on different values of the smoothing scale and the number of quantiles in future work.

We measure the DS autocorrelation and cross-correlation functions of each DS quintile in bins of $\mu$ and $s$ using \textsc{pycorr}, which is a wrapper around a modified version of \textsc{CorrFunc} \citep{corrfunc}. We use $241$ $\mu$ bins from $-1$ to $1$, and radial bins of different widths depending on the scale: 1 Mpc/h bins for $0 < s < 4 \,\, \Mpch$, 3 Mpc/h bins for $4 < s < 30 \,\, \Mpch$, and 5 Mpc/h bins for $30 < s < 150\,\, \Mpch$. Additionally, we measure the galaxy 2PCF adopting these same settings. All the correlation functions are then decomposed into their multipole moments (equation~\ref{eq:multipoles}). In this analysis, we decided to omit the hexadecapole due to its low signal-to-noise ratio, restricting the analysis to the monopole and quadrupole. The multipoles are finally averaged over the three lines of sight. 

Due to the addition of AP distortions, whenever the true cosmology of a mock does not match our fiducial cosmology, the boxes will have non-cubic dimensions while still maintaining the periodicity along the three axes. Both the \textsc{densitysplit} and \textsc{pycorr} codes can handle non-cubic periodic boundary conditions. In the case of \textsc{densitysplit}, we choose to keep the resolution of the rectangular grid fixed, so that $R_{\rm cell} = 5\,h^{-1}{\rm Mpc}$ remains fixed irrespectively of the box dimensions (which, as a consequence, can change the number of cells that are required to span the different boxes). The smoothing scale $R_s$ is also kept fixed to $10 \Mpch$, but since the underlying galaxy positions are AP-distorted, this mimics the scenario we would encounter in observations, where we make a choice of smoothing kernel and apply it to the distorted galaxy overdensity field.

An example of the density split summary statistics for \texttt{c000} and one of the sampled HOD parameters from the latin hypercube is shown in Fig.~\ref{fig:multipoles}.

\subsection{Defining the observable's likelihood}
\label{subsubsec:likelihood}

The data vector for density-split clustering is the concatenation of the monopole and quadrupole of the auto- and cross-correlation functions of quantiles $\Q_0$, $\Q_1$, $\Q_3$, and $\Q_4$. In the case of the galaxy 2PCF, it is simply the concatenation of the monopole and quadrupole. In Appendix~\ref{app:gaussianity}, we show that the likelihood of these data vectors is well approximated by a Gaussian distribution as also demonstrated in \cite{Paillas2022:2209.04310}. We therefore define the log-likelihood as
\begin{align} \label{eq:likelihood}
    \log \mathcal{L}(\mathbf{X}^{\rm obs}|\mathcal{C}, \mathcal{G})  = &\left( \mathbf{X}^{\rm obs} - \mathbf{X}^{\rm theo}\left(\mathcal{C}, \mathcal{G}\right) \right) \nonumber \\
    &{\bf C}^{-1} \left( \mathbf{X}^{\rm obs} - \mathbf{X}^{\rm theo}\left(\mathcal{C}, \mathcal{G}\right) \right)^\top \, ,
\end{align}
where $\mathbf{X}^{\rm obs}$ is the observed data vector, $\mathbf{X}^{\rm theo}$ is the expected theoretical prediction dependent on $\mathcal{C}$, the cosmological parameters, and $\mathcal{G}$, the parameters describing how galaxies populate the cosmic web, referred to as galaxy bias parameters throughout this paper, and $\bf C$ the theoretical covariance of the summary statistics. We will here assume that the covariance matrix is independent of $\mathcal{C}$ and $\mathcal{G}$, and use simulations with varying random seeds to estimate it.  This assumption has been shown to have a neglibible impact in parameter estimation for two-point functions \citep{Kodwani_2019}, although it will need to be revised as the statistical precision of future surveys increases.

In the following section, we demonstrate how we can use neural networks to model the mean relation between cosmological and HOD parameters and the density-split statistics in the generated galaxy mocks.

\subsubsection{Emulating the mean with neural networks}

We split the suite of mocks of different cosmologies (and their corresponding HOD variations) into training, validation and test sets. We assign cosmologies \texttt{c000}, \texttt{c001}, \texttt{c002}, \texttt{c003}, \texttt{c004} and \texttt{c013} to the test set, while $80$ per cent of the remaining cosmologies are randomly assigned to the training and $20$ per cent to the validation set.  See Section~\ref{sec:abacus_summit} for the definition of the different cosmologies.

We construct separate neural-network emulators for the galaxy 2PCF, the DS ACF, and the DS CCF. The inputs to the neural network are the cosmological and HOD parameters, normalized to lie between $0$ and $1$, and the outputs are the concatenated monopole and quadrupole of each correlation function, also normalized to be between $0$ and $1$.
\begin{figure*}
    \centering
    \includegraphics[width=0.8\textwidth]{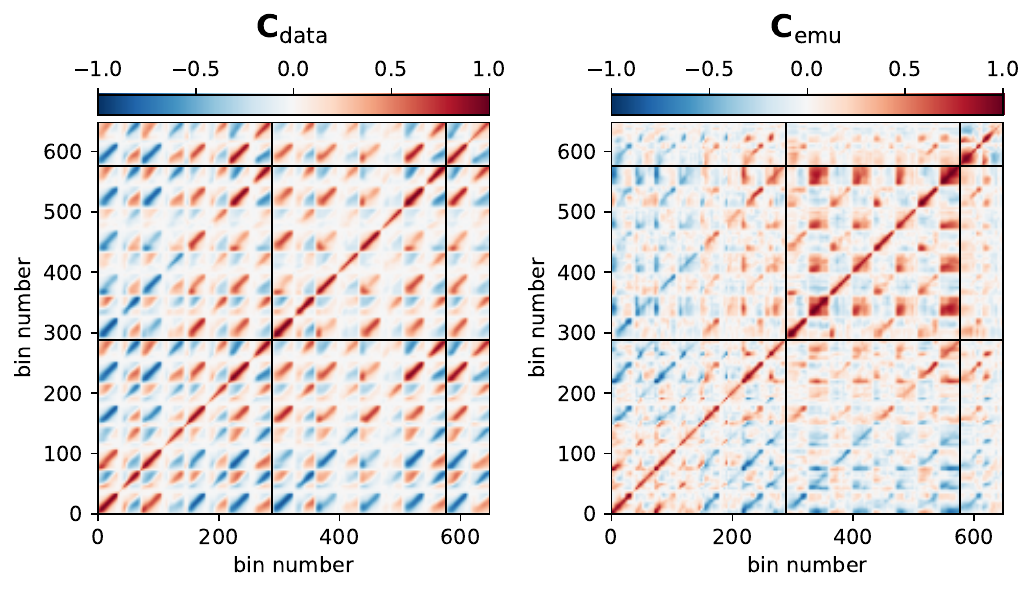}
    \caption{Correlation matrices of the data and model vectors in our clustering analysis. $\bf{C}_{\rm data}$ corresponds to errors associated with the sample variance of the data vector, while $\bf{C}_{\rm emu}$ is associated with the systematic or intrinsic error of the model due to an imperfect emulation. The black horizontal and vertical lines demarcate contributions from the three summary statistics included in the data vector: the density-split cross-correlations and autocorrelation functions, and the galaxy two-point correlation function (listed in the same order as they appear along the diagonal of the correlation matrices).  }
    \label{fig:covariances}
\end{figure*}   
We train fully-connected neural networks with Sigmoid Linear Units as activation functions \citep{elfwing2018sigmoid} and a negative Gaussian log-likelihood as the loss function

\begin{equation}
\label{eq:loss}
\begin{aligned}
&\mathcal{L}(\mathbf{X}|\mu_\mathrm{pred}(\mathcal{C},\mathcal{G}), \sigma_\mathrm{pred}(\mathcal{C},\mathcal{G})) \\
&\quad = \frac{1}{n} \sum_{i=1}^{n} \left( \frac{(X_i - \mu_\mathrm{pred}(\mathcal{C},\mathcal{G}))^2}{2\sigma_\mathrm{pred}(\mathcal{C},\mathcal{G})^2} + \log(\sigma_\mathrm{pred}(\mathcal{C},\mathcal{G}) ^2) + \frac{1}{2} \log(2\pi) \right).
\end{aligned}
\end{equation}

where $\mu_\mathrm{pred}(\mathcal{C},\mathcal{G})$, the mean of the log likelihood, emulates the theory predictions from the N-body simulations, $\sigma_\mathrm{pred}(\mathcal{C},\mathcal{G}) $ models the network's uncertainty in its prediction, and $n$ is the batch size.

We use the AdamW optimisation algorithm to optimise the weights of the neural network, together with a batch size of $256$. In contrast to Adam, AdamW includes L2 regularisation to ensure that large weights are only allowed when they significantly reduce the loss function. To further prevent overfitting, given the limited size of our dataset, we also introduce a dropout factor \citep{JMLR:v15:srivastava14a}. Finally, to improve the model's performance and reduce training time, we decrease the learning rate by a factor of $10$ every $5$ epochs over which the validation loss does not improve, until the minimum learning rate of $10^{-6}$ is reached.

We use optuna\footnote{\url{https://github.com/optuna/optuna}} to find the hyperparameters of the neural network that produce the best validation loss. We optimize the following hyperparameters: learning rate, weight decay controlling the strength of L2 regularization, number of layers, number of hidden units in each layer, and the dropout rate, over $200$ trials. More details related to the neural network architecture and its optimisation can be found on our GitHub repository.\footnote{\url{https://github.com/florpi/sunbird}}

In Section~\ref{sec:validate_emulator}, we present an extensive validation of the emulator's accuracy.

\subsubsection{Estimating the covariance matrix}
\label{sec:covariance}

The likelihood function in equation~\ref{eq:likelihood} requires defining the data vector, expected theoretical mean, and covariance matrix of the summary statistics. The total covariance matrix includes contributions from three sources: i) the intrinsic error of the emulator in reproducing simulations with identical phases to those of the training set ($\bf{C}_{\rm emu}$); ii) the error related to the difference between the fixed-phase simulations used for training and the true ensemble mean (${\bf C}_{\rm sim}$); and iii) the error between the observational data and the mean ($\bf{C}_{\rm data}$),
\begin{equation}
    {\bf C} = {\bf C}_{\rm data} + {\bf C}_{\rm emu} + {\bf C}_{\rm sim}\, .
\end{equation}
 Because the test sample is small and covers a range of cosmologies, to estimate the contribution from the emulator's error to the covariance matrix, we are limited to either assume a diagonal covariance matrix whose diagonal elements are the emulator's predicted uncertainties as a function of cosmological and HOD parameters, $\sigma_\mathrm{pred}(\mathcal{C}, \mathcal{G})$, or we can estimate the emulator error from the test set simulations and ignore its parameter dependence. For the latter, we compute the difference between measurements from the test set and the emulator predictions, $\Delta \mathbf{X} = \mathbf{X}^{\rm emu} - \mathbf{X}^{\rm test}$, and we estimate a covariance matrix as
\begin{equation} \label{eq:cov_emu}
    {\bf C}_{\rm emu} = \frac{1}{n_{\rm test}-1} \sum_{k=1}^{n_{\rm test}}\left( \Delta \mathbf{X}_{k} - \overline{\Delta \mathbf{X}_{k}} \right) \left( \Delta \mathbf{X}_{k} - \overline{\Delta \mathbf{X}_{k}} \right)^\top\, ,
\end{equation}
where the overline denotes the mean across all $600$ test set mocks.

\begin{figure*}
    \centering
        \includegraphics[width=0.95\textwidth]{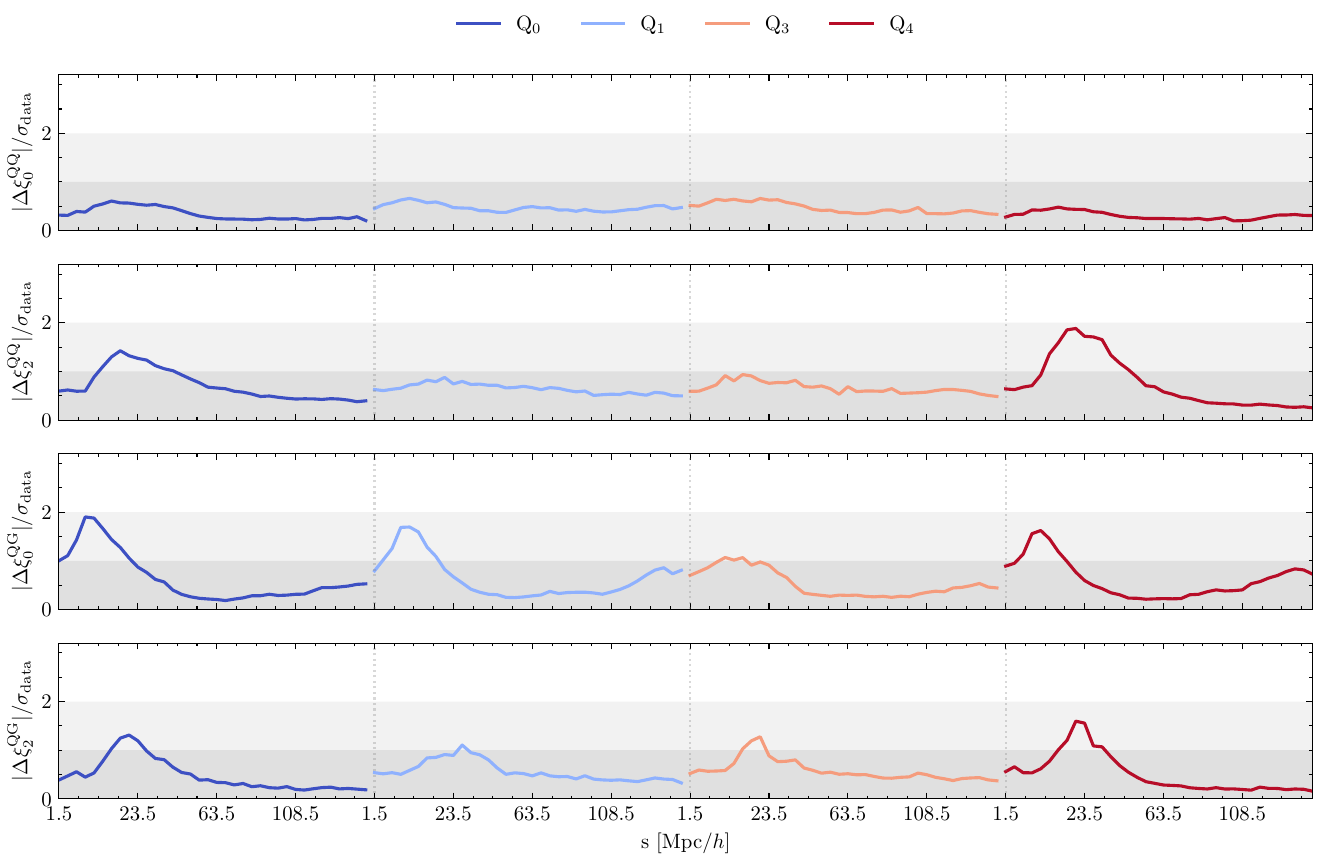}
    \caption{Median absolute emulator errors in units of the data errors, which are estimated for a volume of $2\,h^{-1}\rm Gpc$. We show the monopole ACFs, quadrupole ACFs, monopole CCFs, and quadrupole CCFs in each row. The different density quintiles are shown in different colours.  In Appendix~\ref{ap:fractional_errors}, we show that even though the emulator can be as far as $2 \sigma$ away from the data for the monopole of quintile-galaxy cross-correlations, these are subpercent errors. \href{https://github.com/florpi/sunbird/blob/main/paper_figures/emulator_paper/F3_emulator_errors.py}{\faGithub}} 
    \label{fig:sigma_errors}
\end{figure*}

To estimate ${\bf C}_{\rm sim}$, we do a $\chi^2$ minimisation to choose an HOD catalogue from the fiducial \texttt{c000} cosmology that matches the density-split multipoles measured from BOSS CMASS \citep{Paillas2023:2309.16541}. We then use those HOD parameters to populate dark matter haloes and measure the multipoles from multiple independent realizations of the small \textsc{AbacusSummit} boxes ran with different phases. The covariance is calculated as 
\begin{align} \label{eq:cov_sim}
    {\bf C}_{\rm sim} = \frac{1}{n_{\rm sim} - 1} \sum_{k=1}^{n_{\rm sim}} \left({\mathbf{X}_{k}^{\rm sim}} - \overline{\mathbf{X}^{\rm sim}}\right)\left({\mathbf{X}_{k}^{\rm sim}} - \overline{\mathbf{X}^{\rm sim}}\right)^\top \, .
\end{align}
where $n_{\rm sim} = 1643$. Each of these boxes is $500\,h^{-1}{\rm Mpc}$ on a side, so we rescale the covariance by a factor of $1/64$ to match the $(2\,h^{-1}{\rm Gpc})^3$ volume covered by the base simulations. See \cite{Howlett_2017} for an in depth discussion on rescaling the covariance matrix by volume factors. For a volume such as that of CMASS, the contribution of ${\bf C}_{\rm sim}$ will be almost negligible. However, this will not be true for larger datasets such as those from the upcoming DESI galaxy survey \citep{desi}. Alternatively, the phase correction routine introduced in Appendix B of \cite{Yuan_2022} could be used to reduce this contribution.

The calculation of ${\bf C}_{\rm data}$ depends on the sample that is used to measure the data vector. In this work, we estimate it from multiple realisations of the small \textsc{AbacusSummit} boxes, in the same way as we compute ${\bf C}_{\rm sim}$. Thus, in the current setup, ${\bf C}_{\rm data}={\bf C}_{\rm sim}$. When fitting real observations, however, ${\bf C}_{\rm data}$ would have to be estimated from mocks that match the properties of the specific galaxy sample that is being used, or using other methods such as jackknife resampling. Importantly, the volume of \textsc{AbacusSummit} is much larger than the volume of the CMASS galaxy sample that we are targeting, and therefore we are providing a stringent test of our emulator framework.

In Figure~\ref{fig:covariances}, we show the correlation matrix for both data and emulator. The full data vector, which combines DSC and the galaxy 2PCF, is comprised by 648 bins. This results in covariance matrices with $648^2$ elements, showing significant (anti) correlations between the different components of the data vector. The horizontal and vertical black lines demarcate the contributions from different summary statistics. Starting from the bottom left, the first block along the diagonal represents the multipoles of the DS CCF, for all four quintiles. The second block corresponds to the DS ACF, and the last block corresponds to the galaxy 2PCF. The non-diagonal blocks show the cross-covariance between these different summary statistics.

\section{Validating the neural network emulator}
\label{sec:validate_emulator}
In this section, we present an exhaustive evaluation of the emulator's accuracy by, i) assessing the network's accuracy at reproducing the test set multipoles, ii) ensuring that the emulator recovers unbiased cosmological constraints when the test set is sampled from the same distribution as the training set, iii) testing the ability of the emulator to recover unbiased cosmological constraints when applied to out-of-distribution data. 
\begin{figure*}
    \centering
    \includegraphics[width=0.9\textwidth]{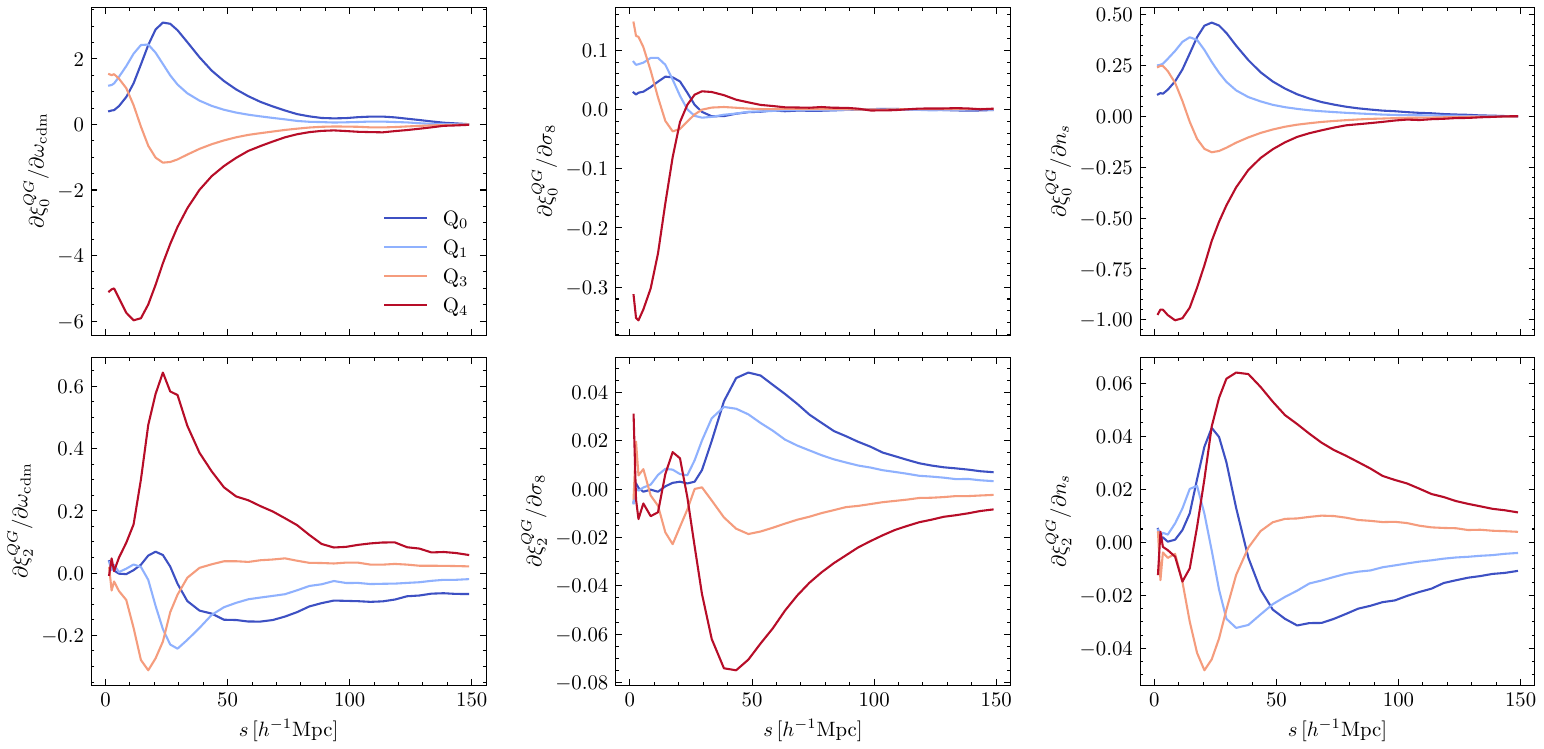}
    \caption{We show the sensitivity of the density split statistics to each cosmological parameter by computing the derivatives of the different quantile-galaxy cross-correlations with respect to the cosmological parameters. From left to right, we show the derivatives with respect to $\omega_\mathrm{cdm}$, $\sigma_8$ and $n_s$, respectively. The upper panel shows the monopole derivatives, whereas the lower panel shows the derivatives of the quadrupole. \href{https://github.com/florpi/sunbird/blob/main/paper_figures/emulator_paper/F4_derivatives.py}{\faGithub}}
    \label{fig:sensitivity_derivatives}
\end{figure*}
\subsection{Testing the accuracy of the emulated multipoles}

We first compare the multipoles measured from the test simulations against the emulator predictions. Figure~\ref{fig:multipoles} shows the density-split and the 2PCF multipoles as measured from one of the HOD catalogues corresponding to the \texttt{c000} cosmology. The HOD catalogue is chosen among the prior samples to maximise the likelihood of the CMASS dataset presented in \cite{Paillas2023:2309.16541}.  The model predictions, which are overplotted as solid lines, show excellent agreement with the data on a wide range of scales. These theory lines are the emulator prediction for the true cosmology and HOD parameters from the mock catalogue.

In the lower sub-panels, we compare the emulator accuracy to the data errors. In this paper, we want to present a stringent test of the emulator and therefore compare its accuracy to that of the \textsc{AbacusSummit} simulations with a volume of $(2\,h^{-1}\rm Gpc)^3$, which is about $8$ times larger than that of the CMASS galaxy sample we are targeting \citep{Paillas2023:2309.16541}. The data errors are estimated from the covariance boxes of the \textsc{AbacusSmall} simulations and are rescaled to represent the expected errors for a volume of $(2\,h^{-1}\rm Gpc)^3$ as explained in Section~\ref{sec:covariance}. In Fig.~\ref{fig:multipoles}, we show that the model prediction is mostly within 1-$\sigma$ of the data vector for this particular example, for both multipoles, and cross-correlations and auto-correlations. 

For a quantitative assessment of the emulator accuracy in predicting multipoles over a range of cosmological parameters, we show in Fig.~\ref{fig:sigma_errors} the median absolute emulator error (taken to be the difference between the prediction and the test data), calculated across the entire test sample, in units of the data errors.  The errors always lie within 2-$\sigma$ of the errors of the data for all scales and summary statistics, and peak at around the smoothing scale.

In Appendix~\ref{ap:fractional_errors}, we show a similar version of this plot where instead of rescaling the vertical axis by the errors of the data, we express everything in terms of the fractional emulator error. While the monopoles of all different density-split summary statistics are accurate within $5\%$, and mostly well within $1 \%$ on small scales, the quadrupoles tend to zero on very small scales, blowing up the fractional error. 

Among all the multipoles, the error is generally larger for the monopole of the DS cross-correlation functions. This is in part due to the sub-percent errors on the data vector below scales of $\sim 40 \Mpch$, but also due to the fact that the sharp transition of the cross-correlation functions below the smoothing scale is overall harder to emulate. The DS autocorrelation emulator errors are almost always within 1-$\sigma$ of the data errors, with the exception of the quadrupole of $\Q_5$. In Fig.~\ref{fig:percent_errors} we see that the emulator accuracy is at subpercent level for the majority of the summary statistics in the analysis.

\subsubsection{Sensitivity to the different cosmological parameters}
After corroborating that the emulator is sufficiently accurate, we explore the dependency of the different summary statistics with respect to the input parameters through the use of derivatives around the fiducial Planck 18 cosmology \citep{Planck2020}.

In Fig.~\ref{fig:sensitivity_derivatives}, we show the derivatives of the quantile-galaxy cross-correlations for the different density environments with respect to the cosmological parameters. In Appendix~\ref{app:derivatives}, we show the corresponding derivatives respect with respect to the HOD parameters, together with those of the quintile autocorrelations. These are estimated by computing the gradient between the emulator's output and its input through \texttt{jax}'s autograd functionality\footnote{\url{https://github.com/google/jax}} which reduces the errors that numerical derivative estimators can introduce.

In the first column of Fig.~\ref{fig:sensitivity_derivatives}, we show that increasing $\omegac$ reduces the amplitude of the cross-correlations for all quantiles, possibly due to lowering the average halo bias. Increasing $\omegac$ also produces shifts in the acoustic peak on large scales for all quantiles. Moreover, the effect on the quadrupole is to reduce its signal for the most extreme quintiles (note that the quadrupole of $\rm Q_0$ is positive, whereas that of $\rm Q_4$ is negative. Note that there are two different RSD effects influencing the quadrupole: on one hand, identifying the density quantiles in redshift space introduces an anisotropy in the quantile distribution, as was shown in \cite{Paillas2022:2209.04310}, and on the other hand, there will be an additional increase in anisotropy in the cross-correlations due to the RSD of the galaxies themselves.

Regarding $\sigma_8$, shown in the second column of Fig.\ref{fig:sensitivity_derivatives}, the effect on the monopoles is much smaller than that on the quadrupole due to enhancing velocities and therefore increasing the anisotropy caused by RSD. 

Finally, the effect of $n_s$ on the monopole is similar to that of $\omega_\mathrm{cdm}$, albeit without the shift at the acoustic scale. Interestingly, the derivative of the quadrupole may change sign near the smoothing scale.

\subsubsection{Evaluating the uncertainty estimates}

While the emulator offers precise mean predictions, its uncertainty estimations present challenges. Specifically, the uncertainty estimates, $\sigma_\mathrm{pred}(\mathcal{C}, \mathcal{G})$, derived from training the emulator to optimize the Gaussian log-likelihood as per Equation~\ref{eq:loss}, tend to underestimate the true uncertainties. This underestimation is problematic as it might introduce biases in our derived cosmological parameter constraints.

\begin{figure}
    \centering
    \includegraphics{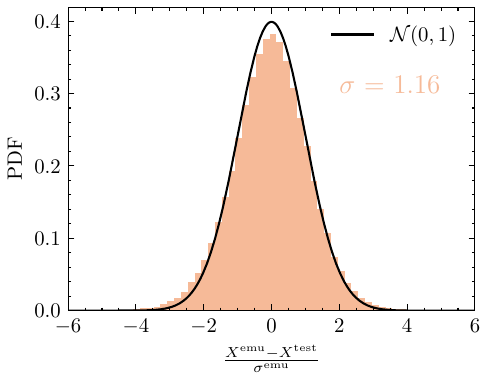}
    \caption{{\bf Z-scores of the emulator unceratinty predictions, compared to a standard normal distribution, $\mathcal{N}(0,1)$, for the test set of the density split cross correlation functions. The emulator predicted uncertainty is over-confident, meaning that tis predicting smaller uncertainties than those observed empirically on the test set. \href{https://github.com/florpi/sunbird/blob/main/paper_figures/emulator_paper/F5_zscores_uncertainties.py}{\faGithub}}}
    \label{fig:zscores}
\end{figure}

To illustrate this, we present the z-score of the emulator's predictions in Figure~\ref{fig:zscores}  for the monopole and quadrupole of the DS CCFs, defined as $z_k = \frac{X_k^\mathrm{emu}-X_k^\mathrm{test}}{\sigma^\mathrm{emu}_k}$. Given that the emulator errors are modeled as Gaussian, the emulator uncertainties would be well calibrated if the distribution of $z_k$'s followed a standard normal distribution. Figure~\ref{fig:zscores} shows that this is not the case, since the z scores show a variance larger than 1 by about a $15\%$. One possible reason for this discrepancy could be the limited size of our dataset. In the remainder of the paper, we will ignore the emulator's predicted uncertainties and quantify its errors by directly estimating them from the test set instead, as described in Equation~\ref{eq:cov_emu}. In the future, we aim to refine the calibration of uncertainty predictions for simulation-based models.  
\subsection{Solving the inverse problem: Recovering the cosmological parameters}
\label{sec:parameter_recovery}
In this section, we focus on the inverse problem, i.e., recovering the mocks' true cosmological parameters from their summary statistics. We will show that the emulator can recover unbiased parameter constraints on the test \textsc{AbacusSummit} HOD catalogues, as well as on a different N-body simulation with a galaxy-halo connection model that is based on another prescription than HOD. We also demonstrate where the density split information comes from by varying various choices of settings in the inference analysis pipeline.

\subsubsection{Recovery tests on \textsc{AbacusSummit}}

\begin{figure*}
    \centering
    \includegraphics[width=0.9\textwidth]{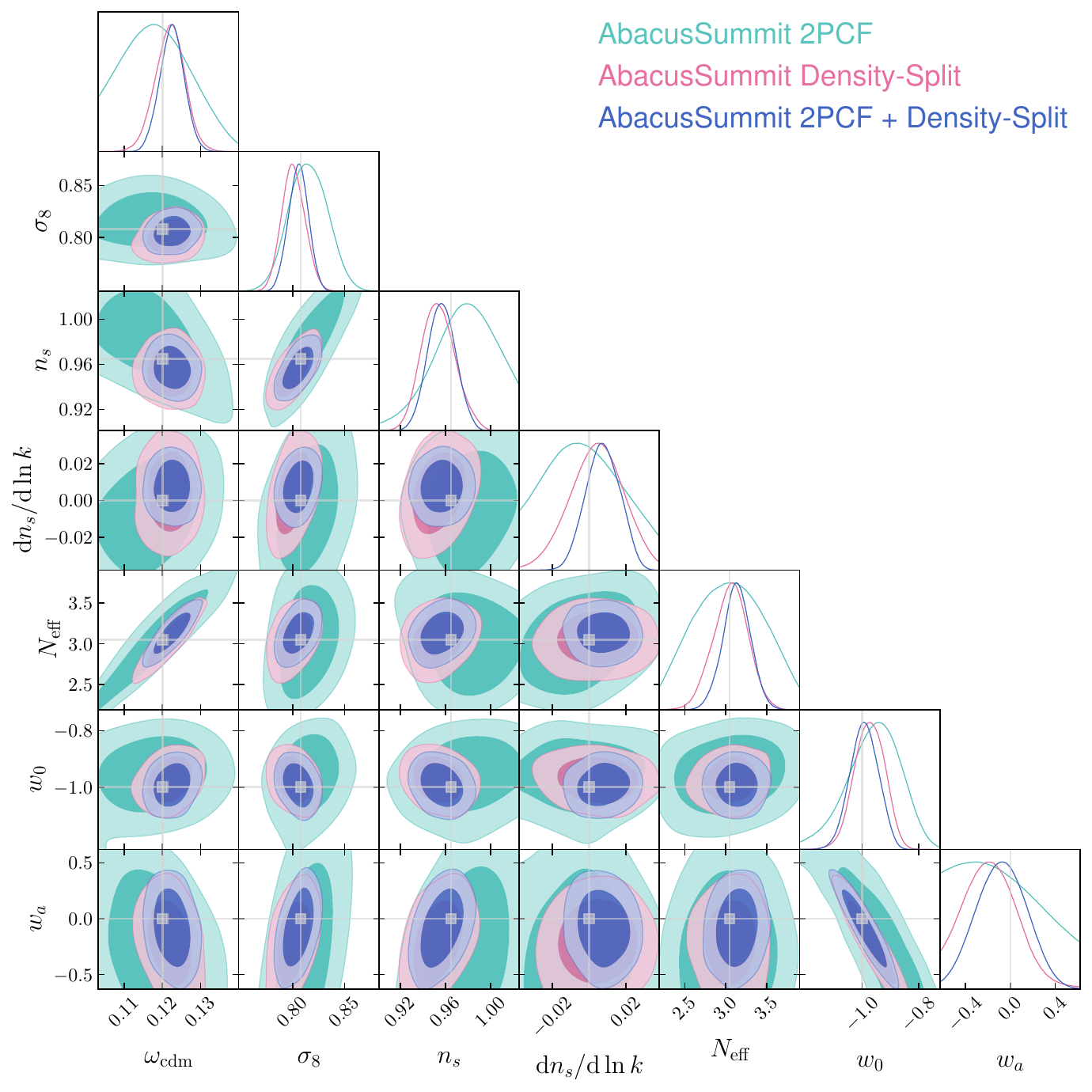}
    \caption{Recovery of \textsc{AbacusSummit} fiducial cosmology (\texttt{c000}) for the set of HOD parameters that minimize the data $\chi^2$ error, after marginalizing over the HOD parameters. We show constraints from the two-point correlation function (2PCF) in green, Density Split statistics (Density-Split) in pink, and a combination of the two (2PCF + Density-Split) in blue. \href{https://github.com/florpi/sunbird/blob/main/paper_figures/emulator_paper/F6_cosmo_inference_c0.py}{\faGithub}}
    \label{fig:full_recovery_fiducial_cosmo}
\end{figure*}
In this section, we show the results of using the emulator to infer the combined set of cosmological and HOD parameters, a total of $17$ parameters, on the test set we reserved from the \textsc{AbacusSummit} simulations, namely those mocks that were not used during the training of the emulator. 

Firstly, for each cosmology from the test set we select the mock catalogue with HOD parameters that maximise the likelihood with respect to a realistic data vector, taken to be the observed density-split multipoles from the BOSS CMASS galaxy sample \citep{Paillas2023:2309.16541}, and infer the posterior of the cosmological and HOD parameters for that particular sample.

Since our model for the mock observables is differentiable, we can take advantage of the estimated derivatives to efficiently sample the posterior distributions through Hamiltonian Monte Carlo (HMC). HMC utilizes the gradient information from differentiable models to guide the sampling process through Hamiltonian dynamics, enabling more efficient exploration of the posterior landscape. It introduces momentum variables and a Hamiltonian function to represent the total energy, then follows the gradients to deterministically evolve the parameters over time while conserving the Hamiltonian. Here, we employ the \textsc{NUTS} sampler implementation from \textsc{numpyro}. We use flat prior ranges for the parameters that match those listed in Table~\ref{tab:prior_emulator}. Fitting one mock takes about one minute on 1 CPU.

\begin{table*}
\label{table:constraints}
\rowcolors{2}{white}{gray!15}
\begin{tabular} {llccc}
\hline
 & Parameter & 2PCF (68\% C.I.) & DSC (68\% C.I.) & 2PCF + DSC (68\% C.I.)\\
\hline
Cosmology & $\omega_{\rm b}$ & $---$ & $0.02257\pm 0.00054$  & $0.02242\pm 0.00050$  \\[0.1cm]
& $\omega_{\rm cdm}$ & $0.1187^{+0.0077}_{-0.010}$ & $0.1220\pm 0.0039$  & $0.1225\pm 0.0032$     \\[0.1cm]
& $\sigma_8$         & $0.815\pm 0.018$ & $0.801\pm 0.011$       & $0.8056\pm 0.0094$                       \\[0.1cm]
& $n_s$              & $0.976^{+0.032}_{-0.023}$ & $0.954^{+0.014}_{-0.016}$   & $0.957\pm 0.012$                 \\[0.1cm]
& $\nrun$         & $-0.003^{+0.018}_{-0.024}$ & $0.004^{+0.015}_{-0.014}$   & $0.0074\pm 0.0090$                  \\[0.1cm]
& $\Neff$       & $3.04\pm 0.40$ & $3.06^{+0.22}_{-0.20}$  & $3.13\pm 0.17$    \\[0.1cm]
& $w_0$              & $-0.959^{+0.10}_{-0.081}$ & $-0.974\pm 0.053$ & $-0.992\pm 0.049$          \\[0.1cm]
& $w_a$              & $< 0.0662$ & $-0.17^{+0.22}_{-0.26}$  & $-0.08\pm 0.22$   \\[0.1cm]
\hline 
HOD & $\log M_1$        & $14.03\pm 0.15$ & $13.94^{+0.17}_{-0.11}$ & $14.01^{+0.12}_{-0.098}$    \\[0.1cm]
& $\log M_{\rm cut}$    & $12.588^{+0.066}_{-0.11}$ & $12.621^{+0.097}_{-0.12}$ & $12.581^{+0.047}_{-0.060}$           \\[0.1cm]
& $\alpha$              & $1.13^{+0.25}_{-0.19}$ & $1.19^{+0.27}_{-0.11}$  & $1.25^{+0.16}_{-0.11}$    \\[0.1cm]
& $\alpha_{\rm vel, c}$ & $0.375^{+0.069}_{-0.054}$ & $0.286^{+0.17}_{-0.089}$ & $0.390^{+0.039}_{-0.033}$ \\[0.1cm]
& $\alpha_{\rm vel, s}$ & $> 1.05$ & $1.08^{+0.18}_{-0.10}$  & $1.09^{+0.11}_{-0.090}$        \\[0.1cm]
& $\log \sigma$         & $-1.54^{+0.98}_{-0.56}$ & $-1.61^{+0.64}_{-0.48}$ & $-1.58^{+0.57}_{-0.50}$       \\[0.1cm]
& $\kappa$              & $---$ & $< 0.830$  & $0.65^{+0.22}_{-0.63}$    \\[0.1cm]
& $B_{\rm cen}$         & $< -0.404$ & $-0.336^{+0.059}_{-0.14}$ & $-0.410^{+0.043}_{-0.060}$          \\[0.1cm]
& $B_{\rm sat}$         & $< -0.0339$ & $-0.11\pm 0.36$   & $-0.37\pm 0.28$  \\[0.1cm]
\hline
\end{tabular}
\caption{Parameter constraints from the galaxy two-point correlation (2PCF), density-split clustering (DSC) and the baseline combination (2PCF + DSC) analyses. Each row shows the parameter name and the corresponding mean and 68 per cent confidence intervals.}
\end{table*}

We first fit \texttt{c000}, the baseline cosmology of \textsc{AbacusSummit}. Figure~\ref{fig:full_recovery_fiducial_cosmo} shows the posterior distribution of the cosmological parameters, marginalised over the HOD parameters. Density split clustering, the galaxy 2PCF, and their combination recover unbiased constraints with the true cosmology of the simulation lying within the 68 per cent confidence region of the marginalised posterior of every parameter. Note that in particular density split statistics contribute to breaking the strong degeneracy between $n_s$ and $\omega_{\rm cdm}$ observed in the 2PCF.

\begin{figure}
    \centering
    \includegraphics[width=\columnwidth]{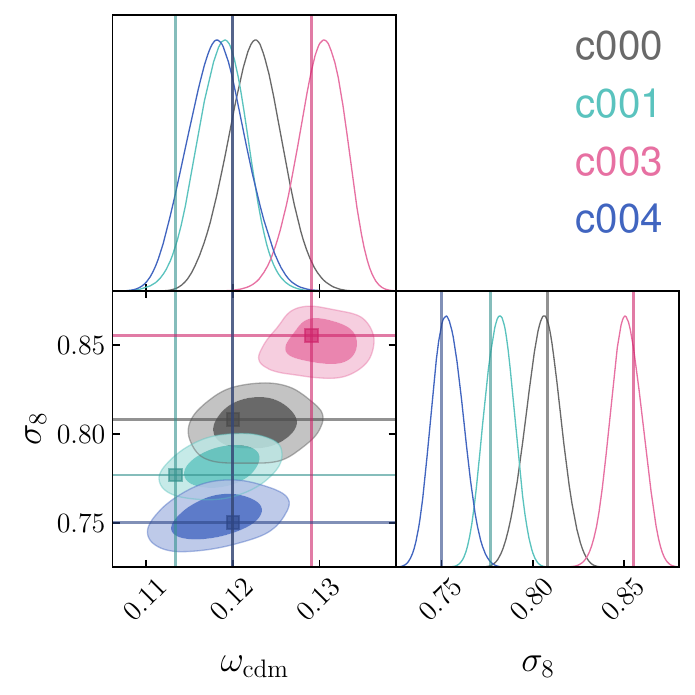}
    \caption{Marginalized constraints from density-split clustering on $\omega_{\rm cdm}$, $\sigma_8$ and $n_s$, derived from fits to mock galaxy catalogues at 4 different cosmologies from our test sample. The true cosmology of each mock is shown by the horizontal and vertical dotted coloured lines. 2D contours show the 68 and 95 per cent confidence regions around the best-fit values. \href{https://github.com/florpi/sunbird/blob/main/paper_figures/emulator_paper/F7_cosmo_inference_c0_c1_c3_c4.py}{\faGithub}}%
    \label{fig:c0vsc4}
\end{figure}
\begin{figure*}
    \centering
    \includegraphics[width=0.95\textwidth]{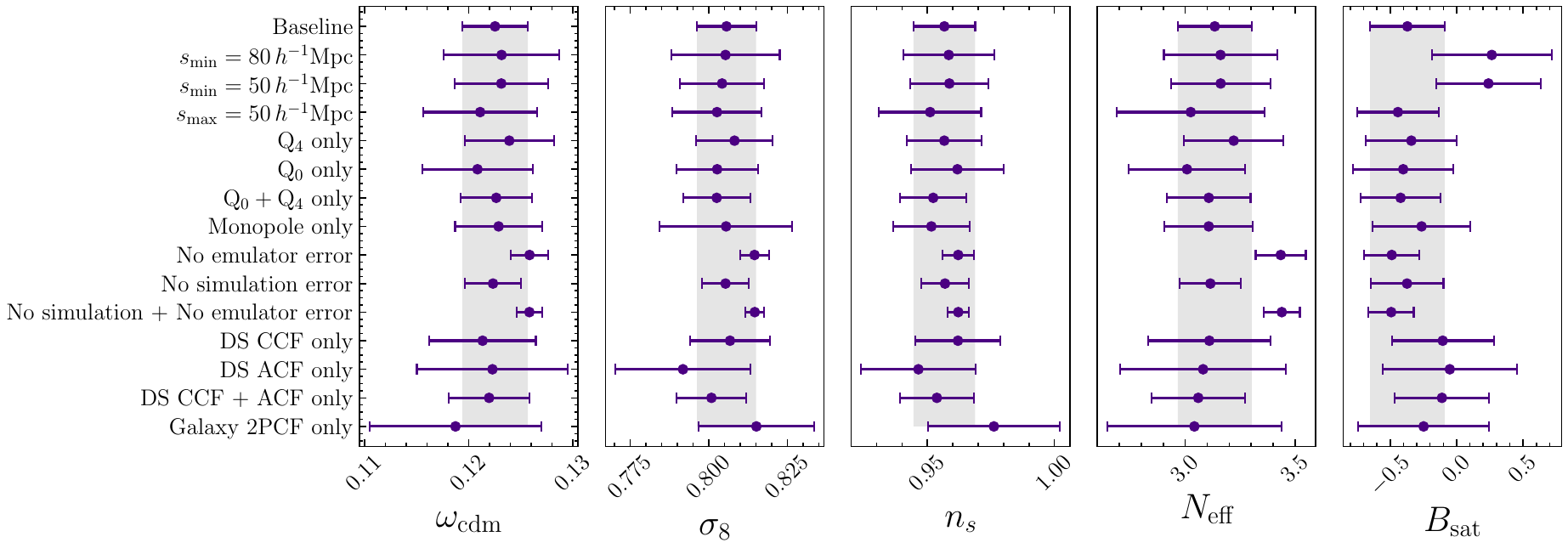}
    \caption{Marginalised constraints on $\omega_{\rm cdm}$, $\sigma_8$, $n_s$, $\Neff$ and $B_{\rm sat}$ for different configurations in the inference analysis. Dots and error bars show the mean and the 68 per cent confidence interval of the parameters, respectively. The uppermost points show the baseline configuration, which consists of the combination of the monopole and quadrupole of the DS cross-correlation and auto-correlation functions for quintiles ${\rm Q_0}$, ${\rm Q_1}$, ${\rm Q_3}$, and ${\rm Q_4}$. \href{https://github.com/florpi/sunbird/blob/main/paper_figures/emulator_paper/F8_whisker.py}{\faGithub}}
    \label{fig:whisker}
\end{figure*}
In Table~\ref{table:constraints}, we show the resulting constraints for each of the three cases tested. For the $(2\, h^{-1}{\rm Gpc})^3$ volume that is considered here, the baseline analysis recovers a  $2.6\%$, $1.2\%$, and $1.2\%$ constraint for $\omega_{\rm cdm}$, $\sigma_8$ and $n_s$, respectively. These constraints are a factor of about $2.9$, $1.9$, and $2.1$ tighter than for the 2PCF, respectively. Moreover, the parameters $\Neff$, and $w_0$ are recovered with a precision of $8\%$, and $4.9\%$ in the baseline analysis. These are in turn a factor of about  $2.5$ and $1.9$ times tighter than for the 2PCF. In an idealised Fisher analysis using simulated dark matter halos \citep{Paillas2022:2209.04310}, we found similar expected improvements for all parameters but $\sigma_8$, for which the Fisher analysis predicted a much larger improvement. 

The posterior distribution of the HOD parameters, marginalised over cosmology, is shown in Figure~\ref{fig:full_recovery_fiducial}. In particular, density split statistics can contribute to significantly tightening the constraints on the environment-based assembly bias parameters, $B_{\rm cen}$ and $B_{\rm sat}$. We expect that reducing the smoothing scale used to estimate densities with future denser datasets would help us attain even tighter constraints on these parameters that may lead to significant detections of the effect in such galaxy samples. Note that for this particular sample some of the true HOD parameters are close to the prior boundary.

Moreover, in Figure~\ref{fig:c0vsc4} we show the marginalised constraints on  $\omega_{\rm cdm}$ and $\sigma_8$ for four particular cosmologies in the test set that vary these two parameters. As before, the HOD parameters are chosen from the prior for each cosmology to maximise the likelihood of CMASS data. These cosmologies are of particular interest since they show that the model can recover lower and higher $\sigma_8$ values than that of the fiducial Planck cosmology. The additional \textsc{AbacusSummit} cosmologies that we are analysing are, \texttt{c001}, based on WMAP9+ACT+SPT LCDM constraints \citep{Calabrese_2017}, \texttt{c003}, a model with extra relativistic density ($\Neff$) taken from the \texttt{base\_nnu\_plikHM\_TT\_lowl\_lowE\_Riess18\_post\_BAO} chain of \citep{Planck2020} which also has both high $\sigma_8$ and $\omega_{\rm cdm}$, and \texttt{c004}, a model with lower amplitude clustering $\sigma_8$.

\subsubsection{Exploring the information content}

In this section, we will delve deep into the effects that removing subsets of the data when analysing the fiducial cosmology \texttt{c000} have on the resulting parameter constraint to analyse what information is being used to constrain each of the parameters. The results are summarised in Figure~\ref{fig:whisker}.

Let us first examine how the constraints vary as a function of the scales included in the analysis. Bear in mind however that we are not truly removing the small scales since the smoothing introduced to estimate densities leaks information from small scales into all the scales. In Figure~\ref{fig:whisker}, we show first the effect of analysing only from the BAO scale, $s_\mathrm{min} = 80 \Mpch$. In that case, we still see significant gains over the full-shape two-point correlation function. For most parameters, however, apart from $n_s$, we find there is more information contained in the smaller scales.

Regarding the different quantiles, most of the information comes from the combination of void-like, $\mathrm{Q}_0$, and cluster-like, $\mathrm{Q}_4$, regions, whereas the intermediate quantiles barely contribute.

Moreover, we have examined the effect of removing the different error contributions on the covariance matrix. Firstly we show that removing the emulator error produces statistically consistent constraints, but about a factor of $2$ tighter for most parameters compared to the baseline. As we will show in the next subsection, our estimated uncertainties are designed to be conservative and therefore removing the emulator error does not lead in this case to extremely biased constraints. In the future, we will work on developing training sets and models that can overcome this limitation and produce more accurate predictions on small scales. This could lead to major improvements on the $\sigma_8$ constraints.

Finally, we demonstrate that cross-correlations between quantiles and galaxies (DS CCF) are on their own the most constraining statistic but there is a significant increase in constraining power obtained when combining them with auto correlations for the parameters $\omega_{\rm cdm}$, $\sigma_8$, and $n_s$.

\subsubsection{Coverage probability test}
We can test the covariance matrix and likelihood using a coverage probability test. Using repeated experiments with true values drawn from the Bayesian prior, we can test that the recovered values have the correct distribution within the likelihood using the chains sampling the posterior \citep{2021arXiv211006581H}. 
 
In simple terms, if you have a $95\%$ confidence interval derived from the likelihood, the expected coverage is $95\%$. That means that, theoretically, we expect that for $100$ repeated trials, the true value should fall within that interval $95$ times. The empirical coverage is what you actually observe when you compare the rank of the true value within the likelihood. Using the same $95\%$ confidence interval, if you applied this method to many samples and found that the true value was within the interval only $90$ times out of $100$, then the empirical coverage for that interval would be $90\%$.

\begin{figure}
    \centering
    \includegraphics{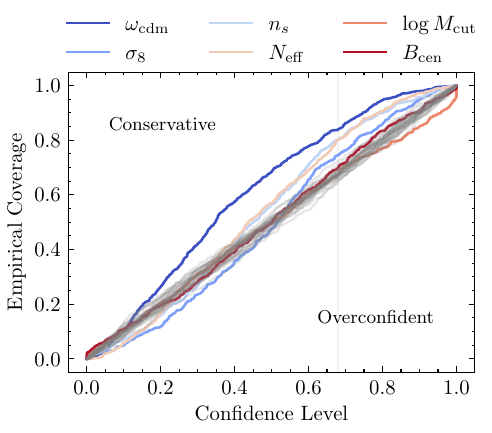}
    \caption{Comparison of the empirical coverage for a given confidence level, shown in different colours for the different cosmological parameters, to the expected coverage, shown in grey. A perfectly calibrated model would follow the one-to-one diagonal. This diagonal has some associated errorbars however, given that we are only using $600$ samples to estimate the coverage, we quantify this by sampling $600$ points from a uniform distribution and estimating its coverage $30$ times. These are the different grey lines plotted in the figure.\href{https://github.com/florpi/sunbird/blob/main/paper_figures/emulator_paper/F9_posterior_coverage.py}{\faGithub}}
    \label{fig:coverage}
\end{figure}

We can use coverage to verify that our covariance estimates are indeed conservative and that we are not subsequently underestimating the uncertainties on the parameters of interest. Note that coverage is simply a measure of the accuracy of the uncertainties, and not of its information content. We estimate the empirical coverage of each parameter on the $600$ test set samples of $p(\theta, X)$, extracted from six different values of the cosmological parameters and $100$ different HOD values for each of them. In Figure~\ref{fig:coverage}, we compare the empirical coverage to the expected one. For a perfectly well-calibrated covariance, all should match up on the diagonal line. A conservative estimator of the covariance and of the likelihood would produce curves above the diagonal, whereas overconfident error estimation would generate curves underneath the diagonal line. Figure~\ref{fig:coverage} shows that we mostly produce conservative confidence intervals from the likelihood, in particular for $\omega_{\rm cdm}$, whereas confidence intervals can be slightly overconfident for $\sigma_8$ although the deviation from the diagonal line is close to the error expected from estimating coverage on a small dataset of only $600$ examples. The HOD parameters are all very well-calibrated.

\subsubsection{Recovery tests on Uchuu}

One of the fundamental validation tests for our emulator is to ensure that we can recover unbiased cosmological constraints when applied to mock catalogues based on a different N-body simulation, and using a different galaxy-halo connection model. The latter is particularly important since the HOD model used to train the emulator makes strong assumptions about how galaxies populate dark matter halos and its flexibility to model the data needs to be demonstrated. 

To this end, we test our model on the Uchuu simulations \citep{Ishiyama2021:2007.14720, Dong-Paez2022:2208.00540,  Oogi2023:2207.14689, Aung2023:2209.12918, Prada2023:2304.11911} and use mock galaxies that were created by \cite{2023Zhai} using subhalo abundance matching \citep[SHAM, e.g.,][]{2006MNRAS.371.1173V, Kravtsov_2004} to populate dark matter haloes with galaxies.  This model assigns galaxies to dark matter halos based on the assumption that the stellar mass or luminosity of a galaxy is correlated with the properties of dark matter halo or subhalo hosting this galaxy. Specifically, we use the method of \cite{2017Lehmann} to assign galaxies to dark matter halos and subhalos. In this method, the property used to rank halos is a combination of the maximum circular velocity within the halo, $v_\mathrm{max}$, and the virial velocity, $v_\mathrm{vir}$. This model also includes a certain amount of galaxy assembly bias, further testing the flexibility of our HOD modeling.

Uchuu is a suite of cosmological N-body simulations that were generated with the GreeM code \citep{Ishiyama2009:0910.0121} at the ATERUI II supercomputer in Japan. The main simulation has a volume of $(2 \Gpch)^3$, following the evolution of 2.1 trillion dark matter particles with a mass resolution of $3.27 \times 10^8 \Msunh$. It is characterized by a fiducial cosmology $\Omegam = 0.3089$, $\Omegab = 0.0486$, $h = 0.6774$, $\sigma_8 = 0.8159$, and $n_s = 0.9667$. Dark matter halos are identified using the \textsc{Rockstar} halo finder \citep{2010ApJ...717..379B}, which is also different from the one implemented in \textsc{AbacusSummit}.

\begin{figure}
    \centering
    \includegraphics[width=\columnwidth]{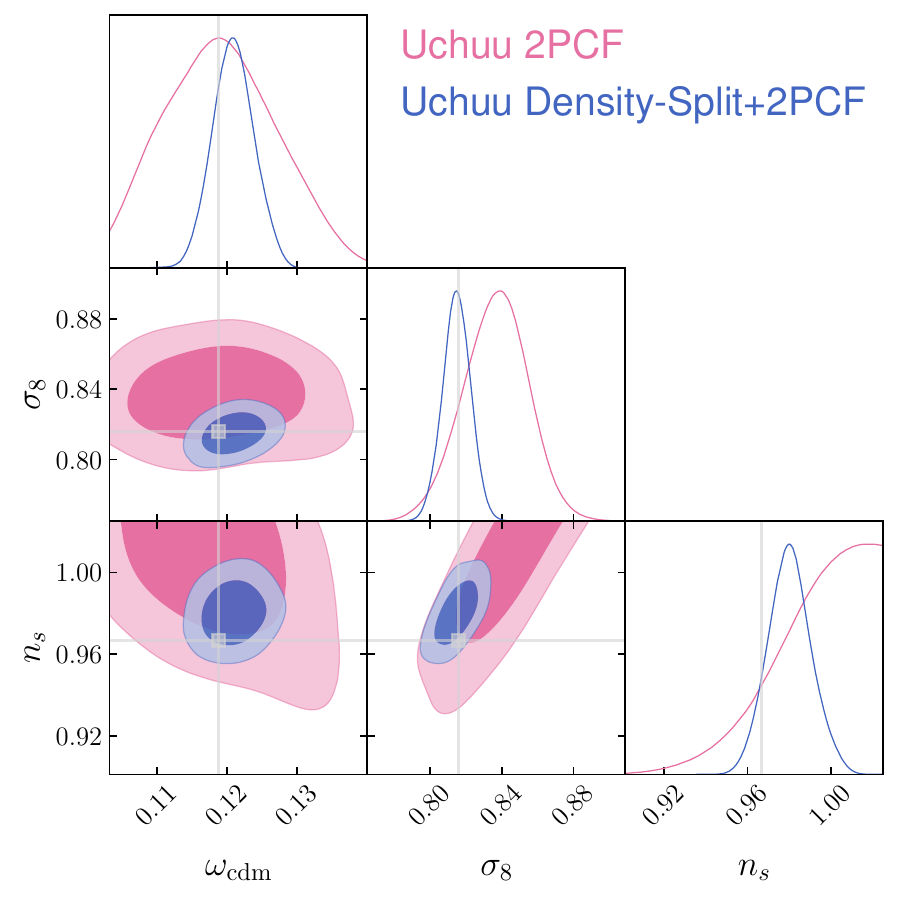}
    \caption{Marginalised posterior on the cosmological parameters when analysing the SHAM mocks based on the Uchuu simulations. We show the contours obtained when analysing only the 2PCF, compared to those found when analysing the combination of the 2PCF and density split statistics. The true parameters that generated the mock are shown in gray. \href{https://github.com/florpi/sunbird/blob/main/paper_figures/emulator_paper/F10_uchuu.py}{\faGithub}}
    \label{fig:uchuu_test}
\end{figure}

Figure~\ref{fig:uchuu_test} shows the resulting marginalised inference using our emulator for both the 2PCF, and the combination of density split with the 2PCF. Note that the constraints on $n_s$ from the 2PCF are in this case completely prior dominated. We can however recover unbiased constraints, even for the stringent test case of a $(2 \Gpch)^3$ volume.

\section{Discussion and conclusions}
\label{sec:discussion}
\subsection{Comparison with previous work}
\subsubsection{Analytical models of density dependent statistics}

Similar definitions of density-split statistics have been presented in \cite{Neyrinck2018, Repp_2021}. In \cite{Neyrinck2018}, the authors defined sliced correlation functions, by slicing the correlation function on local density. They have also presented a model with the Gaussian assumption. In \cite{Repp_2021}, the authors introduce indicator functions by identifying regions of a given density and computing the power spectrum in density bins. This is essentially the Fourier version of the DS ACF. Our analyses have included both the DS CCF and ACF, but finding that the CCF carries most of the cosmological information. These statistics are all similar in spirit.


\subsubsection{Fisher}
In previous work, \cite{Paillas2022:2209.04310} showed with a Fisher analysis the potential of density split statistics to constrain cosmological parameters from dark matter halo statistics. Here, we have confirmed their findings by modelling the density split statistics explicitly as a function of the cosmological parameters, and including the halo-galaxy connection to model the density split statistics of galaxies. 

The improved constraints over two-point correlation functions found here are of a similar magnitude to those in \cite{Paillas2022:2209.04310} for all cosmological parameters, but $\sigma_8$, for which we find weaker constraints. Moreover, we also find that the most extreme quantiles have a similar constraining power and that it is their combination that explains most of the information content of density split statistics. Finally, \cite{Paillas2022:2209.04310} found that density split statistics could break important degeneracies between cosmological parameters that would lead to much tighter constraints on the sum of neutrino masses. This is not something we could corroborate in this paper since variations in neutrino mass are not included in the suite of simulations used in this work, but we plan to work on this in the future by utilising N-body simulations that can accurately simulate the effects of massive neutrinos in the large scale structure \citep{10.1093/mnras/stab2260}.

\subsubsection{Cosmic Voids}
Over the past decade, there has been renewed interest in using cosmic voids to constrain cosmology \citep{pisani2019cosmic}. They have been found to be amongst the most pristine probes of cosmology in terms of how much information is preserved in linear theory at late times \citep{Cai_2016, Hamaus_2017, Nadathur_2018}. However, in practice, extracting cosmological information from voids has proven to be difficult from a purely perturbation theory perspective due to mainly: i) void definitions being difficult to treat analytically and producing different results \citep{Cautun_2018}, ii) identifying voids in redshift space adds additional anisotropy to the observed void-galaxy cross-correlation \citep{Nadathur2019b, Correa_2020}, a similar effect to that found here when estimating densities directly in redshift space, which is difficult to model analytically, and iii) linear theory can only accurately model the mapping from real to redshift space, which means we still require some way of estimating the real space void profiles. In this work, we have shown how emulators can fix all of the above mentioned issues by forward modelling each of these effects.

Moreover, we have shown here how although void-galaxy cross-correlations contain a wealth of information to constrain the cosmological parameters, it is their combination with overdense environments that would give us the tightest constraints.

\subsubsection{The Aemulus emulator}

Related to this work, \citet{storeyfisher2022aemulus} presented an emulation framework based on the \textsc{Aemulus} suite of cosmological N-body simulations to accurately reproduce two-point galaxy clustering, the underdensity probability function and the density-marked correlation function on small scales ($0.1 - 50 \Mpch$). We confirm that including summary statistics beyond two-point functions can improve the cosmological constraints significantly, even after marginalising over the HOD parameters. Moreover, environment-based statistics could lead to a significant detection of assembly bias. As opposed to the marked correlations shown in \cite{storeyfisher2022aemulus}, we estimate densities around random points spread over the survey volume which better samples underdensities in the cosmic web. In the future, it would be interesting to compare the density split constraints to those of density-marked correlation functions, and perhaps the findings in this paper on what environments are most constraining can inform the shape of the mark function used.

\subsection{Conclusions}
\label{sec:conclusions}
We have presented a new simulation-based model for density-split clustering and the galaxy two-point correlation function, based on mock galaxy catalogues from the \textsc{AbacusSummit} suite of simulations. These models allow us to extract information from the full-shape of the correlation functions on a very broad scale range, $1 \Mpch < s < 150 \Mpch$, including redshift-space and Alcock-Paczynski distortions to constrain cosmology and deepen our understanding of how galaxies connect to their host dark matter halos. We have trained neural network surrogate models, or \textit{emulators}, which can generate accurate theory predictions for density-split clustering and the galaxy 2PCF in a fraction of a second within an extended $w\Lambda$CDM parameter space. 

The galaxy-halo connection is modelled through an extended halo occupation distribution (HOD) framework, including a parametrisation for velocity bias and environment-based assembly bias, but the emulator is also validated against simulations that use a subhalo abundance matching framework and a different N-body code to demonstrate the robustness of the method. We have shown that density split statistics can extract information from the non-Gaussian density field that is averaged out in the galaxy two-point correlation function.

Our emulators, which reach a sub-percent level accuracy down to $1 \Mpch$, are able to recover unbiased cosmological constraints when fitted against measurements from simulations of a $(2 \Gpch)^3$ volume. The recovered parameter constraints are robust against choices in the HOD parametrisation and scale cuts, and show consistency between the different clustering summary statistics. 

We find that density-split statistics can increase the constraining power of galaxy 2PCFs by factors of $2.9$, $1.9$, and $2.1$ on the cosmological parameters $\omega_{\rm cdm}$, $\sigma_8$ and $n_s$, respectively. Moreover, the precision on parameters $N_{\rm ur}$, and $w_0$ can be improved by factors of  $2.5$ and $1.9$ with respect to the galaxy 2PCF. Finally, we find density-split statistics to be particularly constraining the environment-based assembly bias parameters. In a companion paper, we show how all these findings result on parameter constraints from the CMASS sample of SDSS \citep{Paillas2023:2309.16541}.

 As we transition to the era of DESI, with its high-density galaxy samples, particularly BGS, alternative summary statistics such as density split have a huge potential to not only increase the precision on cosmological parameter constraints, but to deepen our understanding of how galaxies connect to dark matter haloes. However, this opportunity comes with challenges. The precision that DESI promises requires that our theoretical frameworks are refined to an unprecedented degree of accuracy. It is essential to address these theoretical challenges to fully harness the potential of upcoming observational datasets in cosmological studies.

\section{Acknowledgements}
The authors thank Etienne Burtin for helpful discussions throughout the development of this project. YC acknowledges the support of the Royal Society through a University Research Fellowship. SN acknowledges support from an STFC Ernest Rutherford Fellowship, grant reference ST/T005009/2. FB is a University Research Fellow and has received funding from the European Research Council (ERC) under the European Union’s Horizon 2020 research and innovation program (grantagreement853291). WP acknowledges the support of the Natural Sciences and Engineering Research Council of Canada (NSERC), [funding reference number RGPIN-2019-03908] and from the Canadian Space Agency.

This work is supported by the National Science Foundation under Cooperative Agreement PHY- 2019786 (The NSF AI Institute for Artificial Intelligence and Fundamental Interactions, \url{http://iaifi.org/}). This material is based upon work supported by the U.S. Department of Energy, Office of Science, Office of High Energy Physics of U.S. Department of Energy under grant Contract Number DE-SC0012567, grant DE-SC0013718, and under DE-AC02-76SF00515 to SLAC National Accelerator Laboratory, and by the Kavli Institute for Particle Astrophysics and Cosmology. The computations in this paper were run on the FASRC Cannon cluster supported by the FAS Division of Science Research Computing Group at Harvard University, and on the Narval cluster provided by Compute Ontario (computeontario.ca) and the Digital Research Alliance of Canada (alliancecan.ca). In addition, this work used resources of the National Energy Research Scientific Computing Center (NERSC), a U.S. Department of Energy Office of Science User Facility located at Lawrence Berkeley National Laboratory, operated under Contract No. DE-AC02-05CH11231.

The \textsc{AbacusSummit} simulations were run at the Oak Ridge Leadership Computing Facility, which is a DOE Office of Science User Facility supported under Contract DE-AC05-00OR22725.

We thank Instituto de Astrofisica de Andalucia (IAA-CSIC), Centro de Supercomputacion de Galicia (CESGA) and the Spanish academic and research network (RedIRIS) in Spain for hosting Uchuu DR1, DR2 and DR3 in the Skies \& Universes site for cosmological simulations. The Uchuu simulations were carried out on Aterui II supercomputer at Center for Computational Astrophysics, CfCA, of National Astronomical Observatory of Japan, and the K computer at the RIKEN Advanced Institute for Computational Science. The Uchuu Data Releases efforts have made use of the skun@IAA\_RedIRIS and skun6@IAA computer facilities managed by the IAA-CSIC in Spain (MICINN EU-Feder grant EQC2018-004366-P).

This research used the following software packages: \textsc{Corrfunc} \citep{2020MNRAS.491.3022S}, \textsc{Flax} \citep{flax2020github}, \textsc{getDist} \citep{lewis2019getdist}, \textsc{Jax} \citep{jax2018github}, \textsc{Jupyter} \citep{Kluyver2016jupyter}, \textsc{Matplotlib} \citep{Hunter:2007}, \textsc{Numpy} \citep{harris2020array}, \textsc{numpyro} \citep{phan2019composable}, and \textsc{optuna}.

For the purpose of open access, the authors have applied a CC BY public copyright licence to any Author Accepted Manuscript version arising.

\section*{Data Availability Statement}

The data underlying this article are available in \url{https://abacusnbody.org}.



\bibliographystyle{mnras}
\bibliography{references} 



\appendix

\section{Gaussianity likelihood}
\label{app:gaussianity}

In this section, we check that the likelihood of DS statistics is distributed as multivariate Gaussian, following the analysis in \cite{Friedrich2021}. We first compute the $\chi^2$ value of the summary statistic measured in each of the fiducial simulations
\begin{equation}
    \chi^2_i = \left(\bm{d_i}(\mathbf{s}) - \bm{\bar{d}}(\mathbf{s})\right)^\top C^{-1}  \left(\bm{d_i}(\mathbf{s}) - \bm{\bar{d}}(\mathbf{s})\right),
\end{equation}
where $\bm{d_i}$ represents the value of the summary statistic for the $i$-th fiducial simulation evaluated at the pair separation vector $\mathbf{s}$, $\bm{\bar{d}}(\mathbf{s})$ is the average of the summary statistic over all fiducial simulations at the pair separation vector $\mathbf{s}$, and $C$ is the covariance matrix estimated from all the fiducial simulations.

If the likelihood of the summary statistic is Gaussian distributed, the $\chi^2_i$ values should also follow a $\chi^2$ distribution with degrees of freedom determined by the number of pair-separation bins.

Furthermore, if the likelihood is Gaussian, the distribution of $\chi^2_i$ should also be very close to that of sampling from a multivariate Gaussian with a mean given by $\bar{d}$ and the covariance measured from the simulations.

\begin{figure*}
    \label{fig:gaussianity}
    \centering
        \includegraphics[width=\textwidth]{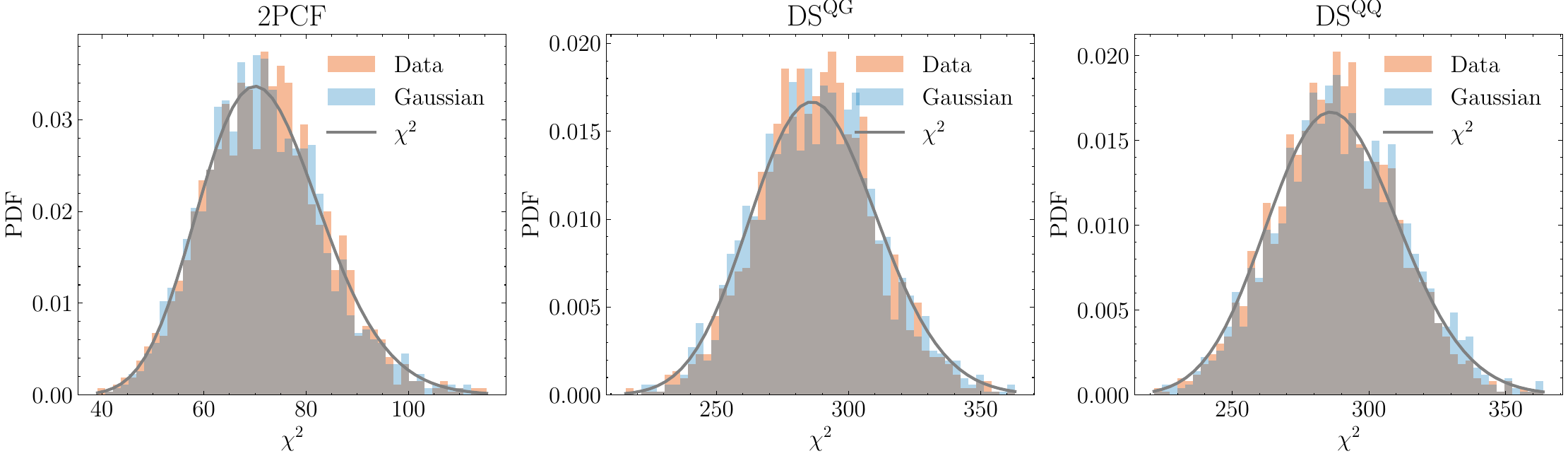}
    \caption{A qualitative assessment of the Gaussianity of the likelihoods for the 2PCF (left), DS galaxy cross-correlations (middle) and DS autocorrelations (right). The colored histograms show the distribution of $\chi^2$ values, as measured from the \textsc{AbacusSmall} simulations (orange) and a multivariate Gaussian distribution with the same mean and covariance as the simulations (blue). The solid line shows a theoretical $\chi^2$ distribution with degrees of freedom set to the number of pair separation bins. \href{https://github.com/florpi/sunbird/blob/main/paper_figures/emulator_paper/A1_gaussian_likelihood.py}{\faGithub}}
\end{figure*}

In Fig.~\ref{fig:gaussianity}, we show how the 2PCF and DS statistics $\chi^2_i$ calculated from the \textsc{AbacusSmall} data follow a very similar $\chi^2$ distribution as that of the random samples generated from a multivariate Gaussian.

\section{Fractional errors}
\label{ap:fractional_errors}
\begin{figure*}
    \label{fig:percent_errors}
    \centering
        \includegraphics[width=0.95\textwidth]{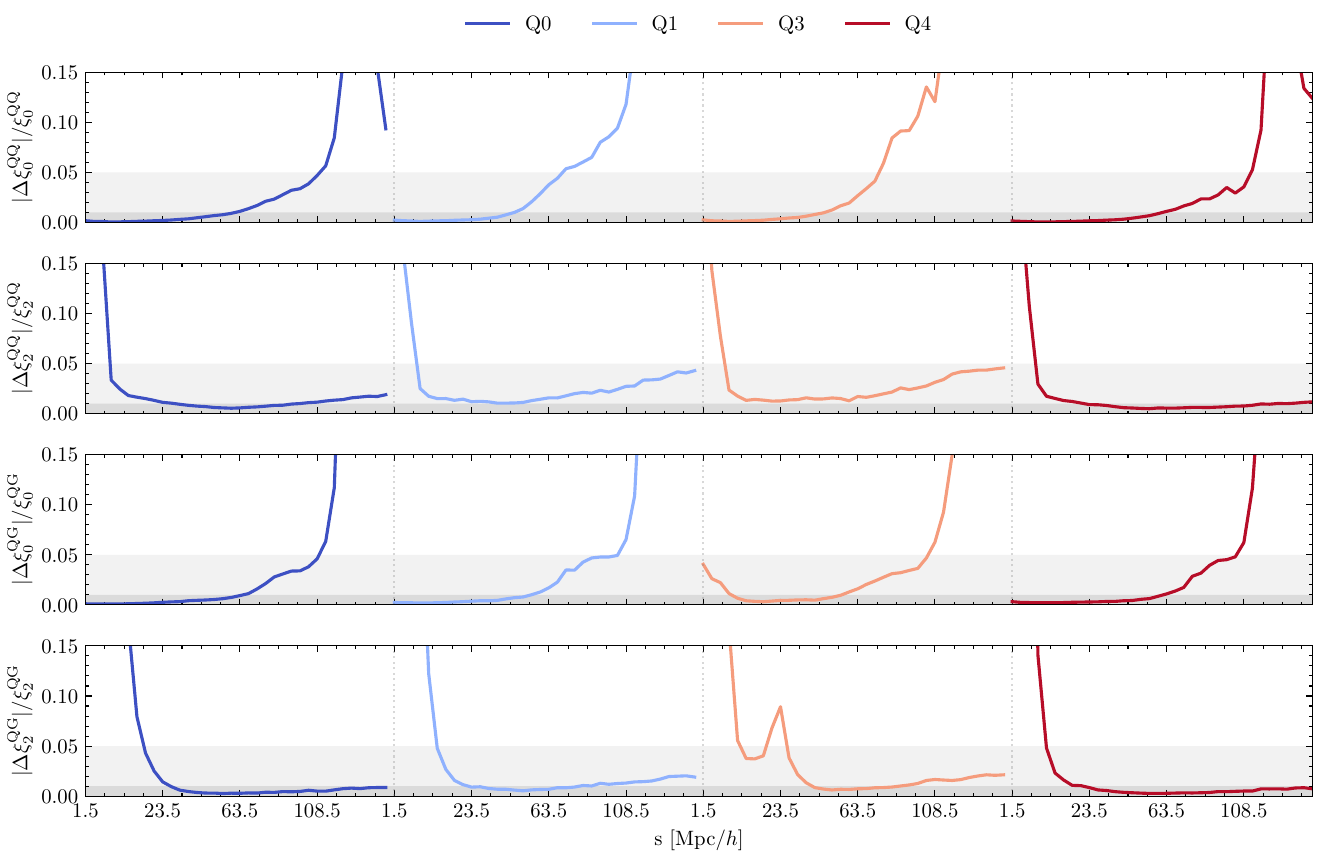}
    \caption{Median absolute fractional errors of the emulator. We show the monopole ACFs, quadrupole ACFs, monopole CCFs, and quadrupole CCFs in each row, estimated from the test set simulations with varying cosmologies and HOD parameters. The different density quintiles are shown in different colours. \href{https://github.com/florpi/sunbird/blob/main/paper_figures/emulator_paper/F3_emulator_errors.py}{\faGithub}}
\end{figure*}

In this appendix, we present the emulator median fractional errors for the different multipoles of each statistic, measured on the test set simulations. Fig.~\ref{fig:percent_errors} shows that the monopoles of all summary statistics are predicted well within $1\%$   for all statistics appart from $\rm Q_4$ cross-correlations, where the errors get closer to $5\%$. Regarding the quadrupole, the fractional errors blow up do to the quadrupole approaching zero on small scales. However, the erorr on large scales is well within $5\%$.

\section{Emulator derivatives respect to the cosmological parameters}
In this section we showcase the cosmological dependence of the different summary statistics by computing the derivative of the statistic respect to the different cosmological and HOD parameters. 

\begin{figure*}
    \centering
    \includegraphics[width=0.9\textwidth]{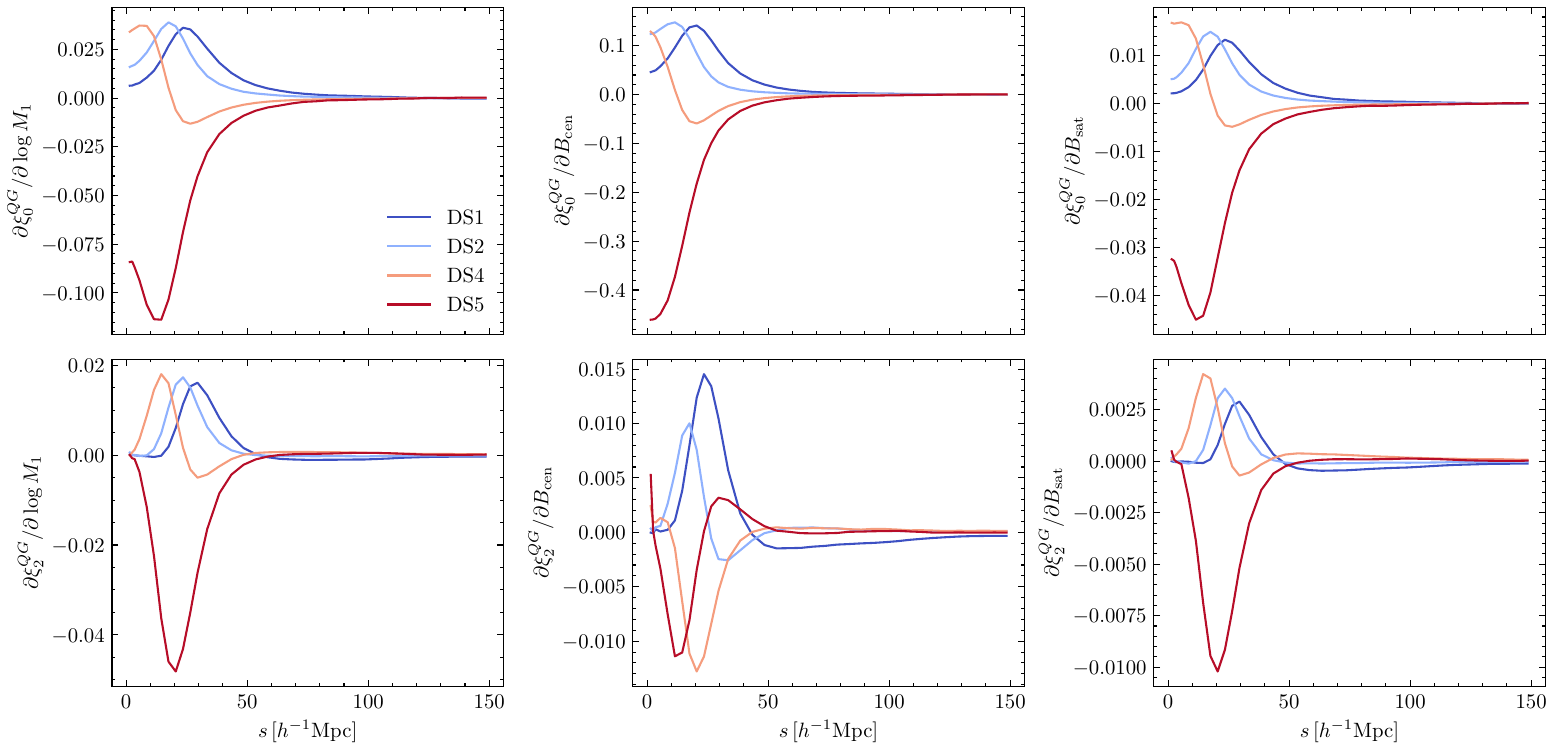}
    \caption{Derivatives of the different quantile-galaxy cross-correlations with respect to the HOD parameters. From left to right, we show the derivatives with respect to $\log M_1$, $B_\mathrm{cen}$ and $B_\mathrm{sat}$, respectively. The upper panel shows the monopole derivatives, whereas the lower panel shows the derivatives of the quadrupole. \href{https://github.com/florpi/sunbird/blob/main/paper_figures/emulator_paper/F4_derivatives.py}{\faGithub}}
    \label{fig:sensitivity_derivatives_cross_hod}
\end{figure*}

In particular, we show the DS CCFs derivatives with respect to different HOD parameters in Figure~\ref{fig:sensitivity_derivatives_cross_hod}. As expected, the impact of the HOD parameters is stronger on small scales.

\begin{figure*}
    \centering
    \includegraphics[width=0.9\textwidth]{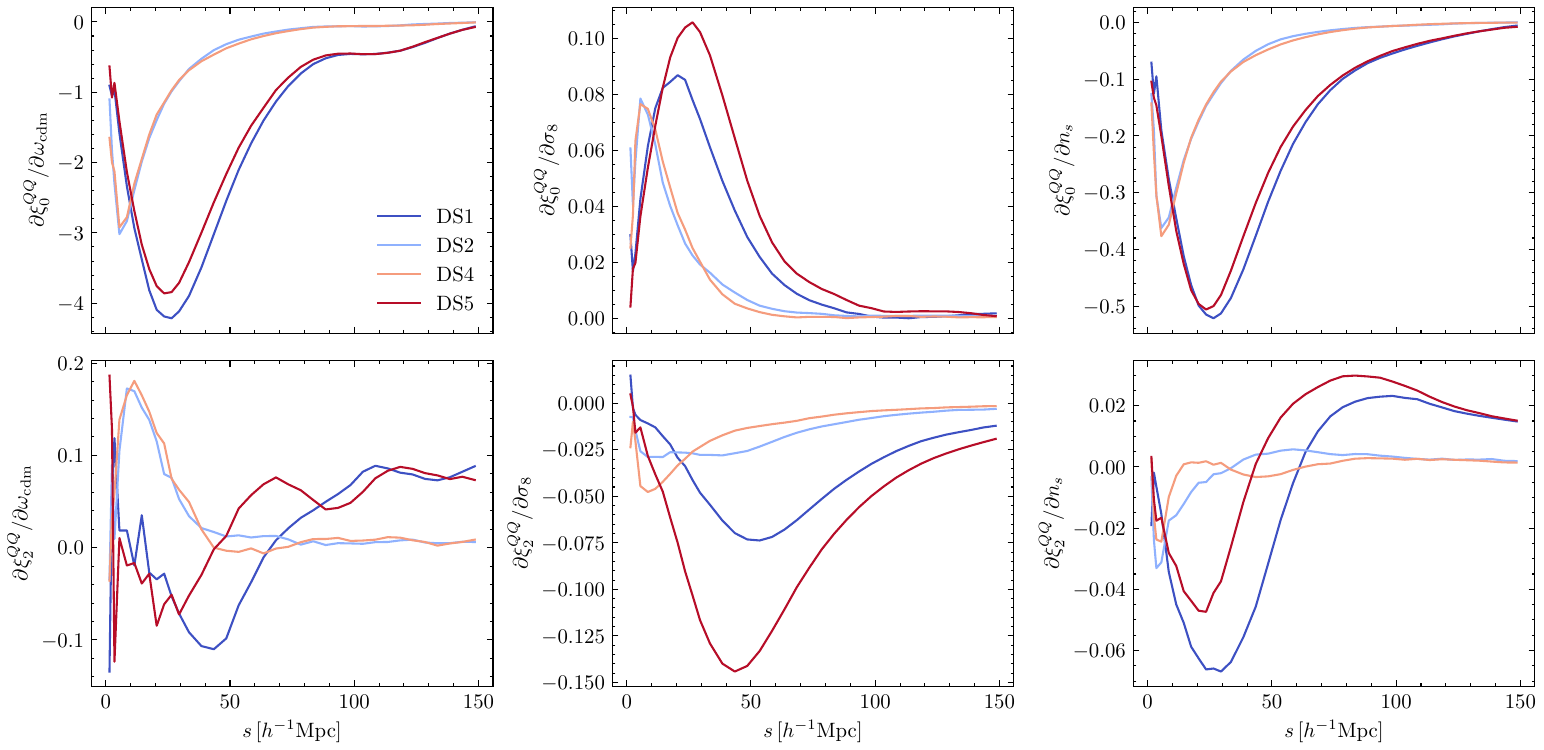}
    \caption{Derivatives of the different density split auto-correlations with respect to the cosmological parameters. From left to right, we show the derivatives with respect to $\omegac$, $\sigma_8$ and $n_s$, respectively. The upper panel shows the monopole derivatives, whereas the lower panel shows the derivatives of the quadrupole. \href{https://github.com/florpi/sunbird/blob/main/paper_figures/emulator_paper/F4_derivatives.py}{\faGithub}}
    \label{fig:sensitivity_derivatives_auto}
\end{figure*}

\begin{figure*}
    \centering
    \includegraphics[width=0.9\textwidth]{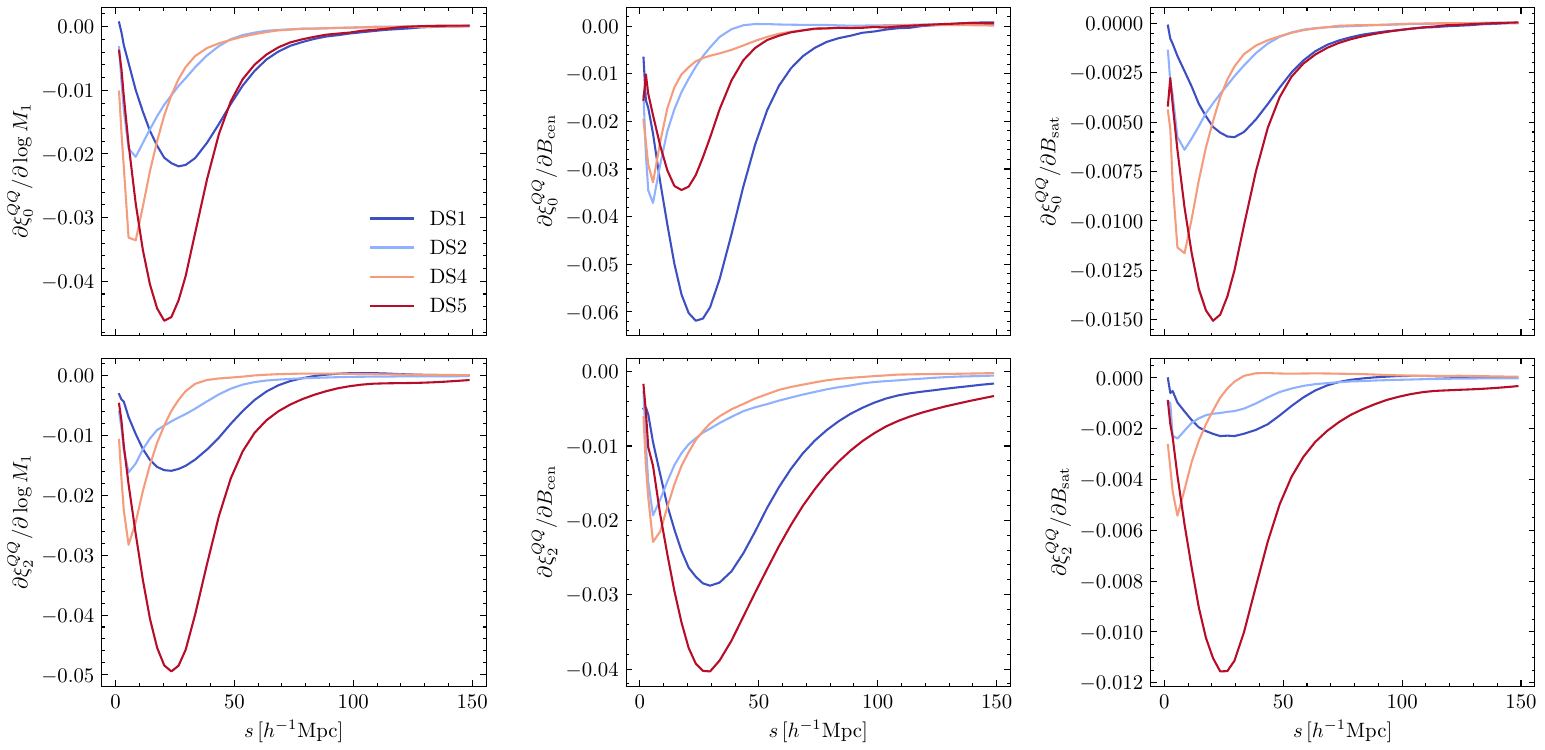}
    \caption{Derivatives of the different density split auto-correlations with respect to the HOD parameters. From left to right, we show the derivatives with respect to $\log M_1$, $B_\mathrm{cen}$ and $B_\mathrm{sat}$, respectively. The upper panel shows the monopole derivatives, whereas the lower panel shows the derivatives of the quadrupole. \href{https://github.com/florpi/sunbird/blob/main/paper_figures/emulator_paper/F4_derivatives.py}{\faGithub}}
    \label{fig:sensitivity_derivatives_auto_hod}
\end{figure*}

In Figure~\ref{fig:sensitivity_derivatives_auto}, we show the derivatives of the DS ACFs respect to the cosmological parameters. As seen on the first pannel, changes in $\omegac$ shift the BAO position of the different density quintiles. Moreover, Figure~\ref{fig:sensitivity_derivatives_auto_hod} shows the derivatives of the same statistic respect to the HOD parameters.

\label{app:derivatives}

\section{Constraints on the HOD parameters}
\begin{figure*}
    \centering
    \includegraphics[width=0.9\textwidth]{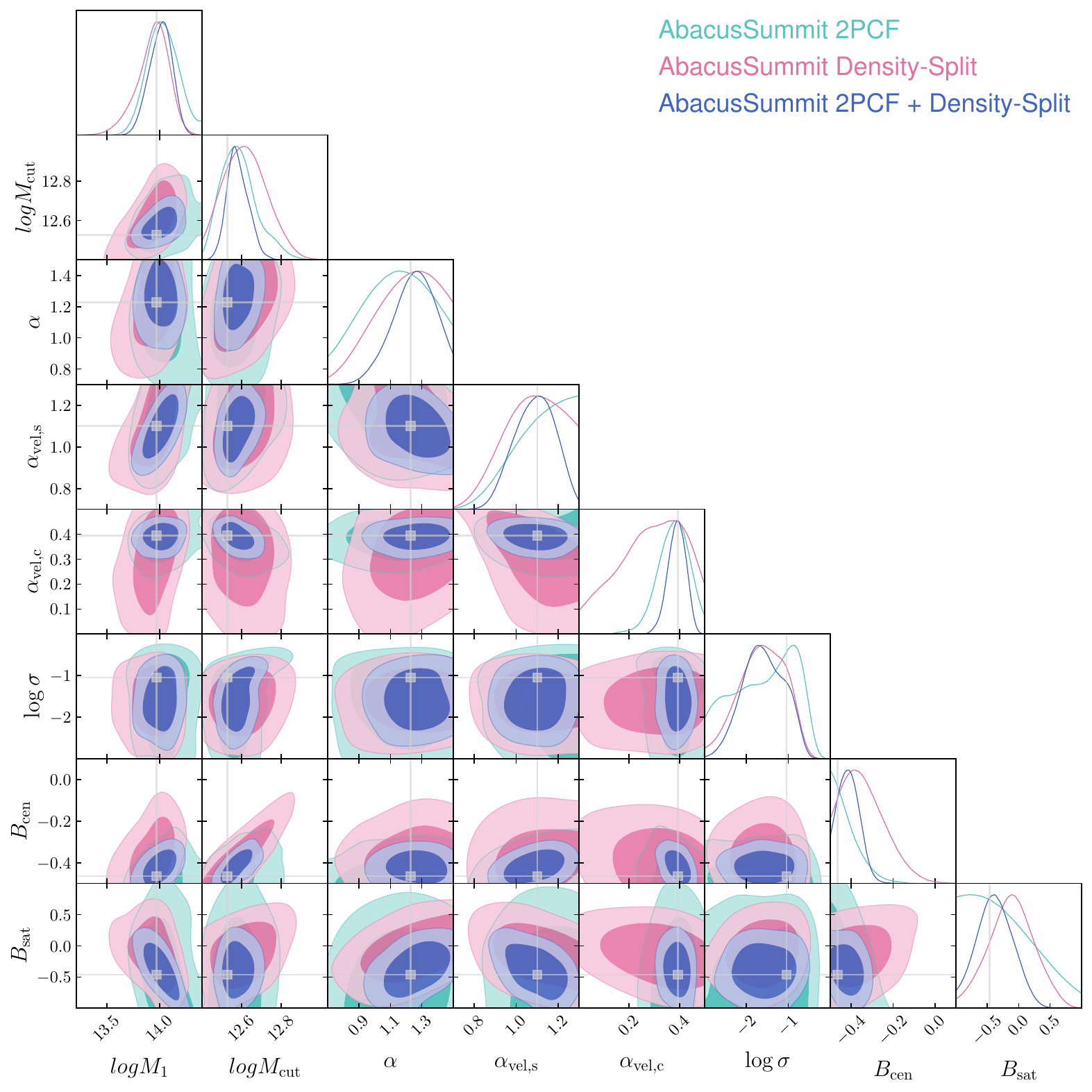}
    \caption{Recovery of the \textsc{AbacusSummit} fiducial cosmology for the set of HOD parameters that minimize the data $\chi^2$ error, after marginalizing over the cosmological parameters. In green, we show the posterior distribution found when only using the galaxy two-point correlation function. In pink, we show those found with density split statistics (CCFs and ACFs). In blue, we show the combination of density split statistics and the two-point correlation function. \href{https://github.com/florpi/sunbird/blob/main/paper_figures/emulator_paper/F6_cosmo_inference_c0.py}{\faGithub}}
    \label{fig:full_recovery_fiducial}
\end{figure*}

Finally, in Figure~\ref{fig:full_recovery_fiducial}, we present the constraints on the HOD parameters obtained by the different summary statistics after marginalising over cosmology, for the \textsc{AbacusSummit} fiducial cosmology. We demonstrate that the combination of 2PCF and density split does indeed recover unbiased constraints.

Although density split does not provide stringent constraints on those parameters that constrain the occupation of satellites (as expected, due to the choice of a large smoothing scale), it can constrain the environment-based assembly bias parameters very accurately in combination with the galaxy 2PCF.
\bsp	
\label{lastpage}
\end{document}